\documentclass[useAMS,senatbib]{mn2e}
\usepackage{graphics}
\def\ltapprox{\raise 2pt \hbox {$<$} \kern-1.1em \lower 5pt \hbox {$\approx$}}
\def\ltsim{\raise 2pt \hbox {$<$} \kern-0.8em \lower 4pt \hbox {$\sim$}}
\def\gtsim{\raise 2pt \hbox {$>$} \kern-0.8em \lower 4pt \hbox {$\sim$}}

\def\ie{{\it i.e.,~}}

\title[Statistics of Giant Radio Halos from
Electron Reacceleration Models]{Statistics of Giant Radio Halos from
Electron Reacceleration Models}

\author[R. Cassano, G. Brunetti, G. Setti]{R. Cassano,$^{1,2}$\thanks{E-mail:
rcassano@ira.inaf.it} G. Brunetti$^{2}$, G. Setti$^{1,2}$\\
$^1$ Dipartimento di Astronomia,Universita' di 
       Bologna, via Ranzani 1, I-40127 Bologna, Italy\\
       $^2$ Istituto di Radioastronomia - INAF, via Gobetti 101,
       I--40129 Bologna, Italy\\}

\begin{document}


\pagerange{\pageref{firstpage}--\pageref{lastpage}} \pubyear{2006}

\maketitle

\label{firstpage}

\begin{abstract}

The most important evidence of non--thermal phenomena
in galaxy clusters comes from Giant Radio Halos (GRHs),
spectacular synchrotron radio sources extended over
$\geq$Mpc scales, detected in the central regions of 
a growing number of massive galaxy clusters. A promising possibility 
to explain these sources is given by {\it in situ} stochastic reacceleration of
relativistic electrons by turbulence generated
in the cluster volume during merger events.
Cassano \& Brunetti (2005) have recently shown that the expected 
fraction of clusters with radio halos and the increase of such a 
fraction with cluster mass can be reconciled with present observations
provided that a fraction of 20-30 \% of the turbulence in clusters is in the
form of compressible modes.

In this work we extend the above mentioned analysis, 
by including a scaling of the magnetic field strength with cluster mass. 
We show that, in the framework of the reacceleration model, 
the observed correlations between the synchrotron radio power 
of a sample of 17 GRHs and the X-ray properties
 of the hosting clusters are consistent with, and actually 
 predicted by a magnetic field dependence on the virial mass
 of the form $B \propto M_v^b$, with $b \gtsim 0.5$ and typical
$\mu$G strengths of the average $B$ intensity.
The occurrence of GRHs as a function of both cluster mass and redshift
is obtained: the evolution of such a probability depends on the interplay between
synchrotron and inverse Compton losses in the emitting volume, and 
it is maximized in clusters for which the two losses are comparable.

\noindent The most relevant findings are that the predicted 
luminosity functions of GRHs are peaked around a 
power $P_{1.4 GHz}\sim 10^{24}$ W/Hz , and severely cut-off at low 
radio powers due to the decrease of the electron reacceleration 
in smaller galaxy clusters, and that the occurrence of GRHs at 1.4 GHz
beyond a redshift $z\sim 0.7$ appears to be negligible. 
As a related check we also show that the predicted integral 
radio source counts within a limited volume
($z \le 0.2$) are consistent with present observational constraints.
Extending the source counts beyond z=0.2 we estimate that the
total number of GRHs to be discovered at $\sim$ mJy radio fluxes  
could be $\sim 100$ at 1.4 GHz. Finally, the occurrence of GRHs
and their number counts at 150 MHz are estimated in view of the fortcoming
operation of low frequency observatories (LOFAR, LWA) and compared with those at higher
radio frequencies.

\end{abstract}


\section{Introduction}

The intracluster medium (ICM) is a mixture of thermal and
non-thermal components and
a precise physical description of the ICM also requires adequate
knowledge of the role of non-thermal components. 

The most detailed
evidence for non-thermal phenomena comes from the radio observations.
A number of clusters of galaxies are known to contain wide diffuse
synchrotron sources (radio halos and relics) which have no
obvious connection with the individual cluster galaxies, but are rather
associated to the ICM (e.g., Giovannini \& Feretti 2000;
Kempner \& Sarazin 2001; see Giovannini \& Feretti 2002 for a review).
The synchrotron emission of such sources
requires a population of GeV relativistic electrons (and/or
positrons) and
cluster magnetic fields on $\mu$G levels.
Evidence for relativistic electrons (and positrons)
in the ICM may also come from the detection
of hard X-ray (HXR) excess emission in the case of a few galaxy clusters
(e.g., Rephaeli \& Gruber  2003, Fusco--Femiano et al. 2004), and possibly from extreme
ultra-violet (EUV) excess emission (e.g., Kaastra et al. 2003;
Bowyer et al. 2004).
It is also believed that the amount of the energy budget of high energy
protons in the ICM might be significant, due to the confinement of cosmic
rays over cosmological time scales
(V\"{o}lk et al. 1996;
Berezinsky, Blasi \& Ptuskin 1997; En\ss lin et al. 1997).
Nevertheless, the gamma radiation that would allow to
infer the fraction of relativistic hadrons in clusters has not been detected
as yet (Reimer et al., 2003, see Pfrommer \& En\ss lin 2004 for upper
limit on this fraction).

Shock waves are unavoidably formed during merger events; 
they may efficiently accelerate relativistic particles
contributing to the injection of relativistic hadrons and of
relativistic emitting electrons in the ICM (e.g., Ryu et al.~2003, Gabici \& Blasi 2003).
However the accelerated electrons have a short pathlength 
due to IC losses and thus they can travel a short distance away from the shock front, emitting synchrotron radiation concentrated around the shock rim
(e.g., Miniati et al. 2001).
Radio Relics, which are polarized and elongated radio sources
located in the cluster peripheral regions, 
may indeed be associated to these shock waves, 
as a result of Fermi-I diffusive shock acceleration of ICM electrons
(En\ss lin et al. 1998; Roettiger et al.~1999), 
or of adiabatic energization of relativistic
electrons confined in fossil radio plasma, released in the past by active
radio galaxies (En\ss lin \& Gopal-Krishna 2001; Hoeft et al. 2004).

The most spectacular evidence of diffuse synchrotron emission
in galaxy clusters is that associated to giant radio halos, Mpc-size 
radio sources which permeate the cluster volume
similarly to the X--ray emitting gas.
In this respect, 
two main possibilities have been investigated in some
detail to explain that GeV electrons (and/or positrons) are present and able to radiate on
distance scales larger than their typical loss lengths: 
{\it i)} the so-called {\it reacceleration} models, whereby 
relativistic electrons (and positrons) injected in the ICM by a
variety of processes active during the life of galaxy clusters 
are continuously re-energized {\it in situ} 
during the life--time of the observed radio halos
(which is estimated to be $\sim 1$ Gyr, Kuo et al.~2004) and {\it ii)}
the {\it secondary electron} models, whereby electrons are secondary 
products of the hadronic interactions of
cosmic rays with the intracluster medium, 
as first proposed by Dennison (1980).
Although the origin of the emitting particles in radio halos is still a matter of
debate (e.g., En\ss lin 2004), the above two models for the production of
the radiating electrons (and positrons) have 
a substantial predictive power,
which can be used to discriminate among such models by comparing their
predictions with observations.
Although future observations remain crucial to 
achieve a firm conclusion, at least as far as the few well
studied clusters and the analysis of statistical samples
are concerned, present
data seem to suggest the presence of {\it in situ}
particle--reacceleration
mechanisms active in the ICM (e.g., Brunetti 2003,04; Blasi 2004;
Feretti et al. 2004; Hwang 2004; Reimer et al. 2004).

Radio observations of galaxy clusters indicate that the detection rate of
radio halos shows an abrupt increase with increasing the 
X-ray luminosity of the host clusters. In particular, about 30-35\% 
of the galaxy clusters with X-ray luminosity larger than $10^{45}$ erg/s  
show diffuse non-thermal radio emission (Giovannini \& Feretti 2002); 
these clusters have also high temperature (kT $>$ 7 keV) and large mass 
($\gtsim$ 2$\times$ $10^{15} M_{\odot}$). 
Furthermore, giant radio halos are always found in merging clusters 
(e.g., Buote 2001; Schuecker et al 2001). 
Although the physics of particle acceleration due to turbulence
generated in merging clusters has been investigated in some detail 
(e.g., Schlickeiser et al.~1987; Petrosian 2001; Fujita et al 2003; 
Brunetti et al.~2004; Brunetti \& Blasi 2005) and the model expectations seem to reproduce
the observed radio features and possibly also the hard X--rays
(e.g., Brunetti et al.~2001; Kuo et al.~2003;
Brunetti 2004; Hwang 2004), a 
theoretical investigation of the statistical properties of
the Mpc diffuse emission in galaxy clusters in the framework of
these models has not been carried out extensively as yet.
In particular, the fact that giant radio halos are always associated to massive
galaxy clusters and the presence of a trend between their radio power
and the mass (temperature, X-ray luminosity)
of the parent clusters may be powerful tools to test and constrain present models.

In a recent paper Cassano \& Brunetti (2005; hereafter CB05) have modelled the 
statistical properties of giant radio halos in the framework of the merger--induced {\it in situ} particle acceleration scenario. By adopting the semi--analytic Press \& Schechter (1974; PS74) theory to follow the cosmic evolution and formation of a large synthetic population of galaxy clusters, it was assumed that
the energy injected in the form of magnetosonic waves during 
merging events in clusters is a fraction, $\eta_t$, of the 
$PdV$ work done by the infalling subclusters in passing through the 
most massive one. Then the processes of stochastic acceleration of the relativistic
electrons by these waves, and the ensuing synchrotron emission properties,
have been worked out under the assumption that the magnetic field intensities,  
ICM temperatures and particle densities (both thermal and non-thermal)
have constant volume averaged values (within 1 Mpc$^3$). The main findings of 
these calculations is that giant radio halos are {\it naturally} expected
only in the more massive clusters, and the expected fraction of clusters with radio halos 
(at redshifts $z\ltsim\,0.2$) can be reconciled with the observed one under viable
assumptions ($\eta_t\,\simeq\,0.24-0.34\,$). Specifically, the probability to form giant radio halos in the synthetic cluster population was found to be of order 20-30 \% in the more massive galaxy clusters ($M > 2\times10^{15}\,M_{\odot}$), 2-5 \% 
in $M \sim 10^{15}\,M_{\odot}$ clusters, and negligible in less massive systems. Such increase of the probability with the cluster mass is essentially due to the increase of both the energy density of turbulence and of the turbulence injection volume with
cluster mass (see CB05).

The present paper is a natural extension of the CB05 work, the most
important difference being that here we adopt a scaling law between 
the rms magnetic field strength (averaged in the synchrotron emitting volume) and
the virial mass of the parent clusters, $B \propto M^b$.
We carry out a detailed comparison between statistical
data of giant radio halos currently available and model expectations as derived by adopting the CB05 procedures.

In Sec.~2 we collect radio and X-ray data for well known
giant radio halos from the literature and derive radio--X-ray correlations.
In Sec.~3 we investigate the possibility to match the observed radio--X-ray 
correlations for giant radio halos with electron acceleration models.
This comparison provides stringent constraints on the physical parameters
in the ICM, in particular for the magnetic field in galaxy clusters. 
In Sec.~4 we derive the expected probability to form giant radio halos 
as a function of $M_v$ and z. This is done by adopting the same values of 
the physical parameters which allows to account for the observed radio--X-ray correlations. In Secs.5--6 we finally calculate the expected luminosity functions 
and number counts of giant radio halos.

As in CB05, we focus our attention on giant radio halos only
(linear size $\sim$1 $h_{50}^{-1}$ Mpc, GRHs elsewhere).
The adopted cosmology is:
$\Lambda$CDM ($H_{o}=70$ Km $s^{-1}$ $Mpc^{-1}$, $\Omega_{o,m}=0.3$,
$\Omega_{\Lambda}=0.7$, $\sigma_8=0.9$).

\begin{figure}
\resizebox{\hsize}{!}{\includegraphics{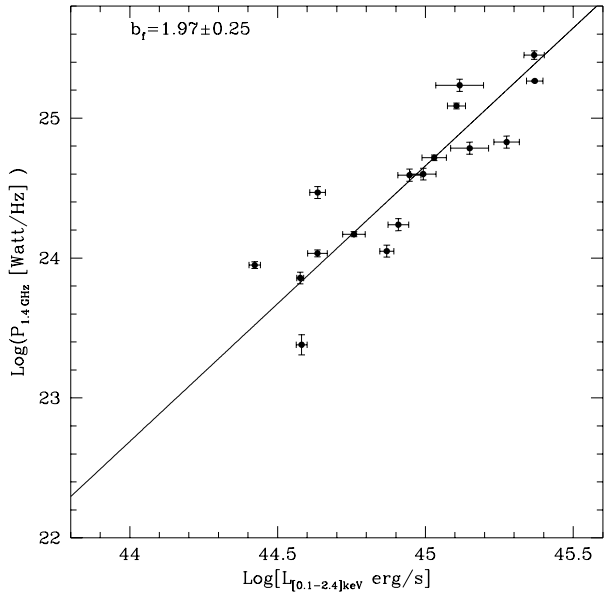}}
\caption[]{Correlation between the radio power at 1.4 GHz and the X-ray 
luminosity between [0.1-2.4] kev for the GRHs.}
\label{LxLrfig}
\end{figure}

\section{Observed Correlations}

In this section we discuss the observed correlations between the X-ray
and the radio properties of clusters hosting GRHs. 

\noindent We collect galaxy clusters with known GRHs 
from the literature obtaining a total sample of 17 clusters. 
In Tab.~\ref{RH} we report the radio and X-ray properties 
of this sample in a $\Lambda$CDM cosmology. In order to have the best estimate 
of the X-ray temperatures we select results from XMM-Newton observations
when available, otherwise we use ASCA results or combine ASCA and Chandra information. 
We investigate the correlations between the X-ray and the radio 
properties of the selected clusters by making use of a linear regression fit 
in log-log space following the procedures given in
Akritas \& Bershady (1996).
This method allows for intrinsic scatter and errors in both variables.

\subsection{Radio Power--X-ray luminosity correlation}

The presence of a correlation between the radio powers and the X-ray
luminosities is well known (Liang et al. 2000; Feretti
2000, 2003; En\ss lin and R\"ottgering 2002).

In Fig.\ref{LxLrfig} we report the correlation between 
the X-ray luminosity (in the 0.1-2.4 keV energy band) and the 
radio power at 1.4 GHz ($P_{1.4}$) for our sample of GRHs. 
The fit has been performed by using the form:

\begin{equation}
\log\Big(\frac{P_{1.4\,GHz}}{3.16\cdot10^{24}\,h_{70}^{-1}\,\frac{Watt}{Hz}}\Big)=A_f+b_f\,
\log\bigg[\frac{L_X}{10^{45}\,h_{70}^{-1}\,\frac{ergs}{s}}\bigg]
\label{LxLreq}
\end{equation}

\noindent where the best fit parameters are: $A_f=0.159\pm 0.060$ and $b_f=1.97\pm 0.25$.

\begin{figure}
\resizebox{\hsize}{!}{\includegraphics{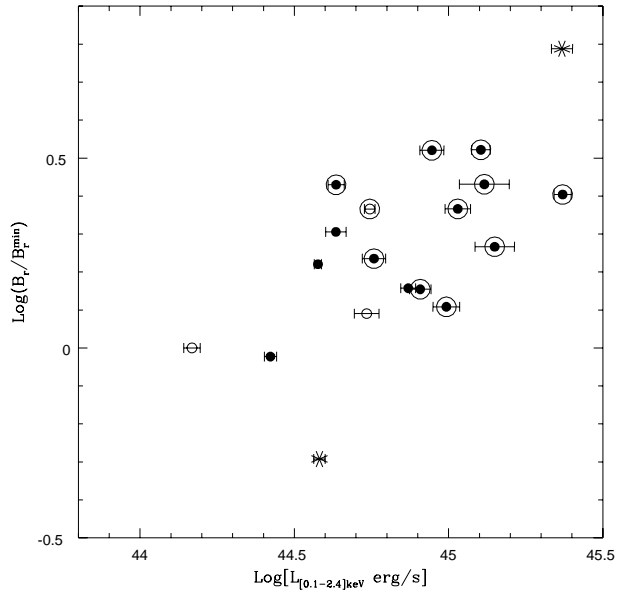}}
\caption[]{Halo radio brightness at 1.4 GHz, 
normalized to the radio brightness of A3562 (which is just visible in the NVSS), 
versus X-ray luminosity between [0.1-2.4] keV. Different symbols 
indicate GRHs (filled circles) and smaller radio halos (open circles)
visible in the NVSS. Asterisks mark A2256, which falls below 
the NVSS surface brightness limit, and 1E50657-558, which falls below the declination
range of the NVSS. Large circles mark objects visible in the NVSS and in the redshift range $z\sim 0.15-0.3$.}
\label{Br}
\end{figure}

\begin{table*}
\caption{Radio and X-ray properties of cluster with GRHs (
linear size $\sim\, 1\, h_{50}^{-1}$ Mpc) in a $\Lambda$CDM cosmology. 
In Col.(1): Cluster name. Col.(2): Cluster redshift. Col.(3): Cluster temperature given in keV.
Col.(4): X-ray luminosity in the energy range $[0.1-2.4]$ keV in unit of $h_{70}^{-2}\,10^{44}$ erg/s.
Col.(5): Bolometric X-ray luminosity in the energy range $[0.01-40]$ keV in unit of $h_{70}^{-2}\,10^{44}$ erg/s.
Col.(6): Radio power at $1.4$ GHz in unit of $h_{70}^{-2}\, 10^{24}$ Watt/Hz. Col.(7): Large Linear Size (LLS) of the Radio Halo is in $h_{70}^{-1}\,kpc$.  
Ref. for the temperature data in brackets: 
(Z04) Zhang al. 2004 (XMM);
(W00) White et al. 2000 (ASCA);
(M96) Markevitch 1996 (ASCA);
(m)   mean value between Mushotzky \& Scharf 1997 (ASCA) and Govoni et al. 2004 (Chandra);
(e)   Ebeling et al. 1996 (from $L_x$-T relation) ;
(D93) David et al. 1993 (Einstein MPC+ Exosat + Ginga);
(M98) Markevitch et al. 1998 (ASCA);
(m1)  mean value between Z04 and Pierre et al. 1999 (ASCA data);
(H93) Hughes et al. 1993 (GINGA).
Ref. for the X-ray luminosities in brackets: 
(B04) Boehringer et al 2004,
(E98) Ebeling et al 1998,
(E96) Ebeling et al 1996, 
(T96) Tsuru et al 1996, 
Ref. for the radio data in brackets:
(L00) Liang et al. 2000 (ATCA)
(F00) Feretti 2000,
(B03) Bacchi et al 2003,
(GF00) Giovannini \& Feretti 2000,
(V03) Venturi et al 2003,
(GFG01) Govoni et al. 2001,
(G05) Govoni et al. 2005,
(FF03) Feretti et al. 2001,
(m2) mean value between Kim et al. 1990 and Deiss et al. 1997}
\begin{tabular}{llllllllllll}
\hline
\hline
cluster's  &  z &   T  & $L_X$ &$L_{bol}$ &   $P_{1.4}$       &  LLS\\
  name     &    & [keV]&  [$10^{44}$ erg/s ]&[$10^{44}$ erg/s ]& [$10^{24}$ Watt/Hz] & [Mpc $h_{70}^{-1}$]\\
\hline
\hline

1E50657-558  & 0.2994  &  $13.59^{+0.71}_{-0.58}$(Z04) &  $23.322\pm 1.84$(B04)&   $88.619\pm 7.00$ &    $28.21\pm 1.97$(L00)  &  1.43\\
A2163        & 0.2030  &  $13.29^{+0.64}_{-0.64}$(W00) &  $23.435\pm 1.50$(B04)&   $82.021\pm 5.24$ &  $18.44\pm0.24$(FF01) &  1.86\\
A2744        & 0.3080  &  ~$\:8.65^{+0.43}_{-0.29}$(Z04)  &  $13.061\pm 2.44$(B04)&   $37.315\pm 6.97$ &   $17.16\pm 1.71$(GFG01)&  1.64\\
A2219        & 0.2280  &  ~$\:9.52^{+0.55}_{-0.40}$(W00)  &  $12.732\pm 0.98$(E98)&   $40.293\pm 4.34$ &  $12.23\pm 0.59$(B03)  &  1.56\\
CL0016+16    & 0.5545  &  ~$\:9.13^{+0.24}_{-0.22}$(W00)  &  $18.829\pm 1.88$(T96)&   $51.626\pm 5.16$ &  ~$\:6.74\pm 0.67$(GF00) &  0.79\\
A1914        & 0.1712  &  $10.53^{+0.51}_{-0.50}$(W00) &  $10.710\pm 1.02$(E96)&   $33.738\pm 3.21$ &  ~$\:5.21\pm 0.24$(B03)  &   1.18\\
A665        & 0.1816  &  ~$\:8.40^{+1.0}_{-1.0}$(M96)    &  ~$\:9.836\pm 0.98$(E98) &   $25.130\pm 3.92$ &  ~$\:3.98\pm 0.39$(GF00) &  1.69\\
A520         & 0.2010  &  ~$\:7.84^{+0.52}_{-0.52}$(m)    &  ~$\:8.830\pm 0.79$(E98) &   $22.841\pm 5.14$ &  ~$\:3.91\pm 0.39$(GFG01)&   1.00\\
A2254        & 0.1780  &  ~$\:7.50^{+0.0}_{-0.0}$(e)      &  ~$\:4.319\pm 0.26$(E96) &   $11.076\pm 0.66$ &  ~$\:2.94\pm 0.29$(GFG01)&  0.86\\
A2256        & 0.0581  &  ~$\:6.90^{+0.11}_{-0.11}$(W00)  &  ~$\:3.814\pm 0.16$(E96) &   ~$\:9.535\pm 0.42$ &  ~$\:0.24\pm 0.02$(F00) &  0.85\\
A773         & 0.2170  &  ~$\:8.39^{+0.42}_{-0.42}$(m)    &  ~$\:8.097\pm 0.65$(E98) &   $21.728\pm 3.62$ &  ~$\:1.73\pm 0.17$(GFG01)&  1.14\\
A545         & 0.1530  &  ~$\:5.50^{+6.20}_{-1.10}$(D93)  &  ~$\:5.732\pm 0.50$(B04) &   $12.608\pm 1.10$ &  ~$\:1.48\pm 0.06$(B03)  &  0.82 \\
A2319        & 0.0559  &  ~$\:8.84^{+0.29}_{-0.24}$(M98)  &  ~$\:7.403\pm 0.41$(E96) &   $20.730\pm 1.14$ &  ~$\:1.12\pm 0.11$(F00) &  1.01\\
A1300        & 0.3071  &  ~$\:9.42^{+0.26}_{-0.25}$(m1)   &  $14.114\pm 2.08$(B04)&   $33.870\pm 4.98$ &  ~$\:6.09\pm 0.61$(F00)  &   0.86\\
A1656        & 0.0231  &  ~$\:8.21^{+0.16}_{-0.16}$(H93)  &  ~$\:3.772\pm 0.10$(E96) &   $10.182\pm 0.26$ & ~$\:0.72^{+0.07}_{-0.04}$ (m2)   &  0.78\\
A2255        & 0.0808  &  ~$\:6.87^{+0.20}_{-0.20}$(W00)  &  ~$\:2.646\pm 0.12$(E96) &   ~$\:6.611\pm 0.30$ &   ~$\:0.89\pm 0.05$(G04)  &  0.88\\
A754         & 0.0535  &  ~$\:9.38^{+0.27}_{-0.27}$(W00)  &  ~$\:4.314\pm 0.33$(E96) &   $12.946\pm 0.98$ &  ~$\:1.08\pm 0.06$(B03)  &  0.96\\
\hline
\hline     
\label{RH}     
\end{tabular}
\end{table*}

Our findings are consistent with those of 
En\ss lin and R\"ottgering (2002) who used 14 clusters with radio halos
and found a correlation of the form 
$P_{1.4\,GHz}\propto L_{X}^{1.94}$.
Using 16 clusters with GRHs Feretti (2003) found a correlation
between the X-ray bolometric luminosity 
and the radio power at 1.4 GHz of the 
form $P_{1.4\,GHz}\propto (L_{X}^{bol})^{1.8\pm0.2}$.
A consistent result is obtained 
with the data in Tab.~1 ($P_{1.4\,GHz} \propto (L_{X}^{bol})^
{1.74\pm 0.21}$).

\begin{table*}
\caption{Parameters of the $\beta$-fit and cluster mass estimated for the 16 galaxy 
clusters with GRHs for which $\beta$-fits are avaiable. Col.(1): Cluster name. Col.(2): $\beta$-parameter value
with 1$\sigma$ error. Col.(3): Core radius in units of $h_{70}^{-1}$ kpc and corresponding uncertainty.
Col.(4): Virial mass and is uncertainty in units of $h_{70}^{-1}\,10^{15}\,M_{\odot}$.
Col.(5): Virial radius in units of $h_{70}^{-1}$ kpc. Col.(6): Mass estimated inside the core radius 
in units of $h_{70}^{-1}\,10^{13}\,M_{\odot}$. Ref. for the (data) source in brackets: (a) Markevitch et. al 2002 (Chandra); 
(b) RB02 (ROSAT for $\beta$-fit and T as in table 1);
(c) Govoni et al. 2001 (ROSAT); (d) Ettori \& Fabian 1999 (ROSAT); 
(e) Ettori et. al 2004 (Chandra);
(f) Feretti 2004 (Einstein); (g) Lemonon et al. 1997 (ROSAT).}
\begin{tabular}{llllll}
\hline
\hline
cluster's     &  $\beta$&  $r_c$           & $M_v$ & $R_v$ & $M_c$  \\
name          &      & [kpc $h_{70}^{-1}$] & [$10^{15}$ $M_{\odot} $] & [kpc $h_{70}^{-1}$]  & [$10^{13}$ $M_{\odot} $]   \\
\hline
\hline
1E50657-558(a)&    $0.70\pm 0.07$      &  $179\pm 18$  & $3.43\pm 0.38$ & 3301 &
~$\:9.50\pm  1.40 $ \\
A2163      (b)&    $0.80\pm 0.03$      &  $371\pm 21$  & $4.32\pm 0.26$ & 3766 & 
$ 22.00\pm 1.84$ \\
A2744      (c)&    $1.00\pm 0.08$      &  $458\pm 46$  & $2.87\pm 0.26$ & 3096 & 
$ 22.10\pm 2.96$ \\
A2219      (d)&    $0.79\pm 0.08$     &  $343\pm 34$  & $2.52\pm 0.28$ & 3104 & 
$ 14.40\pm 2.16$ \\
CL0016+16  (e)&    $0.68\pm 0.01$     &  $237\pm 80$   & $1.47\pm 0.05$ & 2166 &
~$\:8.27\pm 0.38 $\\
A1914      (b)&    $0.75\pm 0.02$     &  $165\pm 80$   & $2.90 \pm 0.15$ & 3356 & 
~$\:7.28\pm 0.51$ \\
A665       (f)&    $0.74\pm 0.07$     &  $350\pm 35$  & $1.97\pm 0.30$ & 2933 & 
$ 12.10\pm 2.20$ \\
A520       (c)&    $0.87\pm 0.08$     &  $382\pm 50$  & $2.22\pm 0.25$ & 3018 & 
$ 14.50\pm 2.51$ \\
A2256      (b)&    $0.91\pm 0.05$     &  $419\pm 28$  & $2.23\pm 0.13$ & 3281 & 
$ 14.70\pm 1.28$ \\
A773       (c)&    $0.63\pm 0.07$     &  $160\pm 27$    & $1.52\pm 0.19$ & 2636 & 
~$\:4.72\pm 0.98$ \\
A545       (d)&    $0.82\pm 0.08$     &  $286\pm 29$  & $1.25\pm 0.84$ & 2562 & 
~$\:7.20\pm 4.89$   \\
A2319      (b)&    $0.59\pm 0.01$     &  $204\pm 10$  & $1.71\pm 0.07$ & 3009 & 
~$\:5.95\pm 0.38$  \\
A1300      (g)&    $0.64\pm 0.01$     &  $171\pm 80$     & $1.71\pm 0.06$ & 2609 & 
~$\:5.76\pm 0.33$  \\
A1656      (b)&    $0.65\pm 0.02$     &  $246\pm 15$  & $1.83\pm 0.07$ & 3136 & 
~$\:7.38\pm 0.53$   \\
A2255      (b)&    $0.80\pm 0.05$      &  $419\pm 28$  & $1.76\pm 0.12$ & 2996 & 
$ 12.80\pm 1.22$  \\
A754       (b)&    $0.70\pm 0.03$      &  $171\pm 12$  & $2.42\pm 0.11$ & 3379 & 
~$\:6.25\pm 0.52$    \\
\hline
 \hline
\label{beta_fit}
\end{tabular}
\end{table*}

\begin{figure*}
\resizebox{\hsize}{!}{\includegraphics{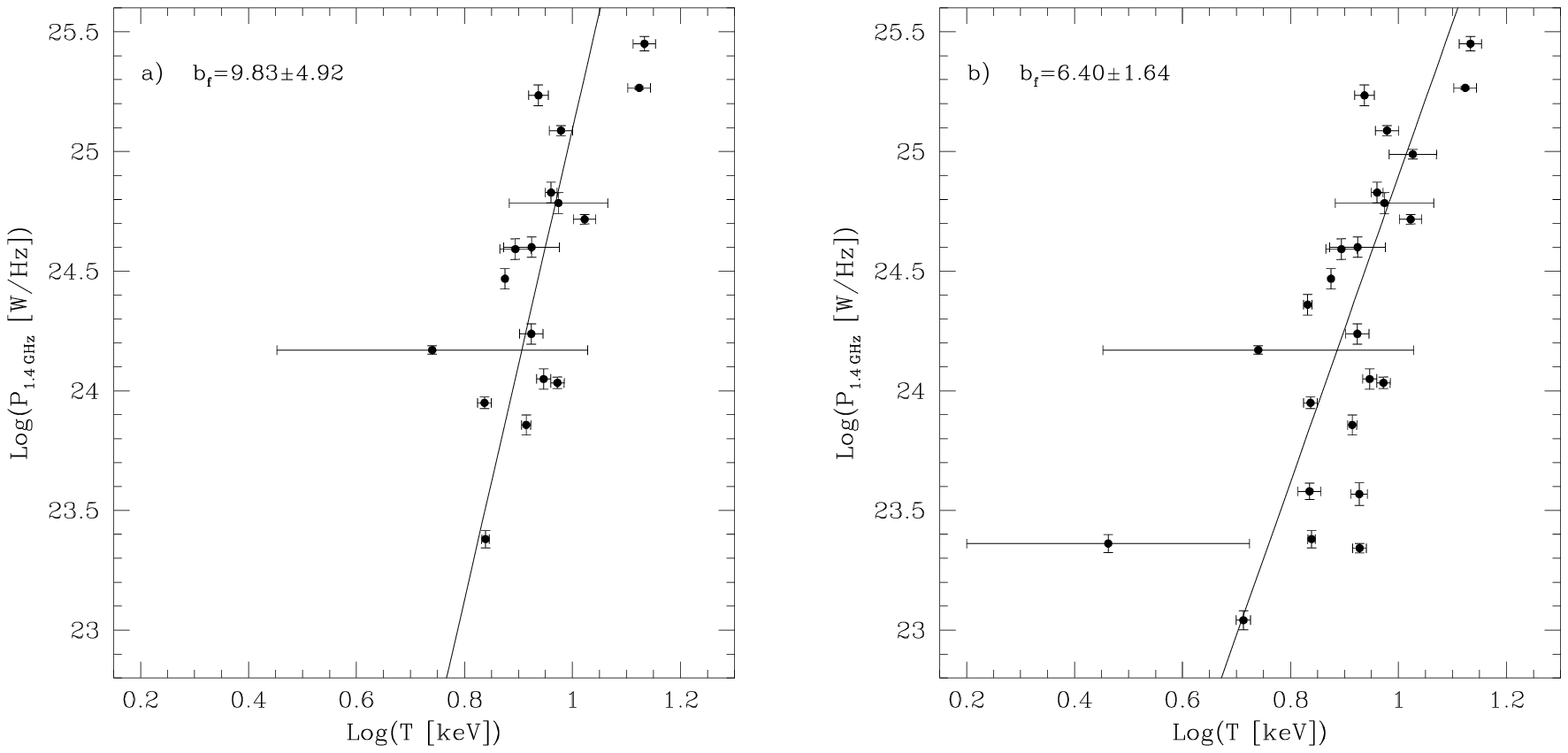}}
\caption[]{Panel a):correlation between the radio power at 1.4 GHz and the 
temperature for the GRHs; Panel b): correlation between the radio 
power at 1.4 GHz and the X-ray temperature for a total sample of 24 cluster 
with a GRHs or with a smaller size ($\sim 200-700$ kpc $h_{50}^{-1}$).}
\label{LrTfig}
\end{figure*}

Although the trend in Fig.\ref{LxLrfig} appears quite stringent,
one may wonder wether it could be affected by observational biases. 
It should be pointed out that most GRHs have been discovered 
by follow ups of radio halo candidates mostly
identified from the NRAO VLA Sky Survey (NVSS, 
Condon et al. 1998). 
 
\noindent In Fig.\ref{Br} we plot the average 
radio surface brightness of GRHs at 1.4 GHz, normalized to the radio brightness 
of A3562 (open circle with the smallest X-ray luminosity), 
which is just visible in the NVSS, versus the X-ray luminosity 
of the hosting clusters. We note that all GRHs have a radio surface  
brightness which is well above that of A3562 (with the exception of A2256 not 
visible in the NVSS). The fact that clusters in the redshift
range $z\sim 0.15-0.3$ have similar radio brightnesses indicates
that the correlation in Fig.\ref{LxLrfig} is not driven by the radio
surface limit. Indeed, these clusters have $L_X\sim 3\cdot 10^{44}-
 3\cdot 10^{45}$ erg/s and range over more than one order of magnitude
 in radio power, whereas the effect of 
 brightness dimming, due to the small z-range, is limited to within 
 a factor $\sim 1.6$. 
 We also note the presence of a trend, with the average radio brightness 
 increasing with X-ray luminosity (see also Feretti 2004), 
which further supports the notion that the correlation
in Fig.\ref{LxLrfig} at high luminosities ($L_X\,\gtsim\; 10^{45}$ erg/s) 
is not driven by selection effects. Furthermore, relatively deep 
upper limits for non radio-halo clusters are now available 
and in some cases lie well below (a factor of $\gtsim\,10$) 
the trend in Fig.\ref{LxLrfig} at X-ray luminosities $\gtsim\,5\cdot10^{44}$ erg/s 
(Dolag 2006).

On the other hand one may argue that the NVSS tends to select
only the most powerful GRHs associated with $L_X\, \ltsim\; 5\cdot10^{44}$ erg/s
clusters. To evaluate the effect of a possible bias at these luminosities,
we perform a fit by considering only GRH clusters with 
$L_X\, \gtsim\, 5\cdot10^{44}$ and find a slope $2.22\pm 0.36$ 
which is consistent within 1 $\sigma$ with Eq.~\ref{LxLreq}.

Thus we conclude that despite the poor statistics, the derived
correlations (Eq.\ref{LxLrfig}) stands on sound observational basis.

\subsection{Radio Power--ICM temperature correlation}

We also investigate the correlation between the radio power at 1.4 GHz 
and the X-ray ICM temperature.
A $P_{1.4}-T$ correlation was first noted by Liang et al.(1999)
and Colafrancesco (1999); with a sample of only 8 radio halos 
the last author obtained a steep trend of the form $P_{1.4}\propto T^{6.25^{+6.25}_{-2.08}}$.
In Fig.~\ref{LrTfig}a we report the best fit for our sample. 
The fit has been performed using the form:

\begin{equation}  
\log\bigg[\frac{P_{1.4\,GHz}}{3.16\cdot10^{24}\,h_{70}^{-1}\,\frac{Watt}{Hz}}\bigg]=A_f+b_f\,\log\Big(\frac{T}{8 \,keV}\Big)
\label{TLreq}
\end{equation}

\noindent and best fit parameters are: $A_f=-0.390\pm 0.139$ 
and $b_f=9.83\pm 4.92$. We note that the observed $P_{1.4}-T$
correlation is very steep, it seems rather a "wall" than a correlation
and it is dominated by the large errors of the cluster
temperatures avaiable to date. 
In order to test the strength of this correlation we try to increase
the sample by including also 7 additional clusters with smaller 
(size $\sim 200-700$ kpc $h_{50}^{-1}$) radio halos. 
In Fig.~\ref{LrTfig}b we report the best-fit $P_{1.4}-T$
obtained for the extended sample, which has a slope $b_f=6.40\pm 1.64$. 
Given the large uncertainties 
we note that the two correlations are consistent
at the 1$\sigma$ level and, in addition, 
the value of the 
lower allowed bound of the two slopes is almost the same.

\subsection{Radio Power -- virial mass correlation}

\begin{figure}
\resizebox{\hsize}{!}{\includegraphics{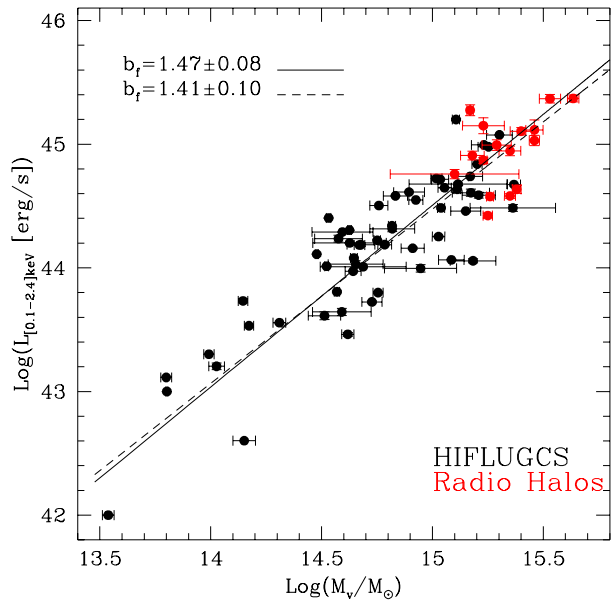}}
\caption[]{Correlation between the X-ray luminosity [0.1-2.4] keV and 
the virial cluster mass: for the HIFLUGCS sample (black points) plus 
the 16 clusters with GRHs (red points, excluding A2254 for
which no information on the $\beta$-model are available) (solid line) and
for the HIFLUGCS sample alone (dashed line).}
\label{LMfig}
\end{figure}

The most important correlation for our study is that between the 
virial mass ($M_v$) of a cluster and the radio power at 1.4 GHz.
This correlation is indeed extensively 
used in the calculations of the RHLFs and number counts 
(Sec.7) and in constraining the values of the magnetic field
in galaxy clusters to be used in our calculations (Sec.3). 
On the other hand, this is also the most difficult correlation to derive 
since it is very difficult to measure the cluster mass.
Govoni et al. (2001) first obtained a correlation between the radio
power and cluster gravitational mass (within 3 $h_{50}^{-1}$ Mpc radius) 
estimated from the surface brightness profile of the X-ray image using 
6 radio halo clusters. This correlation was confirmed by Feretti (2003) 
who extended the sample to 10 cluster radio halos and obtained a best fit 
of the form $P_{1.4}\propto M^{2.3}$, where $M$ is, again, the gravitational mass 
computed within 3 $h_{50}^{-1}$ Mpc from the cluster center. 
However it should be pointed out that while the X-ray mass determination 
method gives good results in relaxed clusters, it may fail in the case of merging
clusters (e.g., Evrard et al. 1996, Roettiger et al. 1996; Schindler 1996). 
This is because the merger may cause substantial deviation from hydrostic equilibrium and spherical symmetry. As a result, the masses in merging clusters can be either overestimated (up to twice the true mass in the presence of shocks) or underestimated (since substructures tend to flatten the average density 
profile giving an underestimation of the order of 50 \% of the true mass; see Schindler 2002). 
In addition, if the temperature systematically decreases with increasing radius, 
then the isothermal assumption leads to an overestimation of the cluster mass of
about 30\% at about six core radii (Markevitch et al. 1998). 

The effect of the scattering produced by all these uncertainties can hopefully 
be reduced by making use of large cluster samples. 
Thus, we choose to obtain the $P_{1.4\,GHz}-M_v$ correlation by combining 
the $L_x-M_v$ correlation, obtained for a large statistical sample of galaxy clusters, with the $P_{1.4}-L_x$ correlation previously derived (Eq.\ref{LxLreq}, Fig.~\ref{LxLrfig}). We use a complete sample of 
the X-ray--brightest clusters (HIFLUGCS, the Highest X-ray FLUx Galaxy Cluster Sample) 
compiled by Reiprich \& B$\ddot{o}$hringer (2002) (hereafter RB02). 
We use this sample of luminous clusters ($L_x \sim 10^{44}-10^{45}$erg s$^{-1}$)
since it is large and homogeneously studied. It consists of 63 bright clusters 
with galactic latitude $|b_{II}|>20^{o}$, flux $f_X(0.1-2.4\, keV)\ge 2\times 10^{-11}$ 
ergs s$^{-1}$ cm$^{-2}$ and it covers about 2/3 of the whole sky.

\noindent The clusters have been reanalyzed 
in detail by RB02 using mainly ROSAT PSPC pointed observations. 
RB02 report the value of $\beta$ and 
core radius, $r_c$, for all the 63 clusters
obtained by fitting the surface brightness profile of the X-ray image with a standard
$\beta$-model. Then under the assumption that the intracluster gas is in hydrostatic 
equilibrium and isothermal (using the ideal gas equations), the gravitational cluster 
mass within a radius r is given by  (e.g., Sarazin 1986):

\begin{equation}
M_{tot}(<r)=\frac{3K_{B}T r^{3}\beta}{\mu m_p G}\left(\frac{1}{r^2+r_c^2}\right),
\label{mteq}
\end{equation}

\noindent Eq.\ref{mteq} gives the total (dark matter plus gas) mass of 
the cluster as a function of radius; then one must define a physically meaningful radius to compare masses of different clusters. It is frequently convenient to use
$r_{200}$ meaning the radius within which the mean total mass density is 200 times the critical density of the universe at the cluster redshift. This is because $M_{200}$, 
the mass contained within $r_{200}$, is usually 
taken as a good approximation of the virial mass since in the spherical collapse 
model the ratio between the average density within the virial 
radius and the mean cosmic density at redshift z is $\Delta_{c}=18\pi^2\simeq 178$ independent of the redshift for $\Omega_m=1$ (e.g., Lacey \& Cole 1993). 
In general, the value of $\Delta_{c}$ depends on the adopted cosmology.
In the $\Lambda$CDM cosmology $\Delta_{c}$ is given by (Kitayama \& Suto 1996):

\begin{equation}
\Delta_{c}(z)=
18\pi^2(1+0.4093\omega(z)^{0.9052}),
\label{Dc}
\end{equation}

\noindent 
where $\omega(z)\equiv \Omega_{f}(z)^{-1}-1$ with:

\begin{equation}
\Omega_{f}(z)
=\frac{\Omega_{m,0}(1+z)^3}{\Omega_{m,0}(1+z)^3+\Omega_{\Lambda}},
\label{Omz}
\end{equation}

\noindent Thus, by using Eq.\ref{mteq} we calculate the virial 
radius, $R_v$, as the radius at which the ratio between the average density
in the cluster and the mean cosmic density at the redshift of the cluster
is given by $\Delta_c(z)$ (Eq.\ref{Dc}). The virial mass, $M_v$, 
and the virial 
radius are thus related by:

\begin{equation}
R_{v}=\Bigg[\frac{3M_{v}}{4\pi\Delta_{c}(z)\rho_{m}(z)}
\Bigg]^{1/3}
\label{Rv}
\end{equation}

\noindent where $\rho_{m}(z)=2.78\times10^{11}\,\Omega_{m,o}\,(1+z)^3\,h^2\,M_{\odot} Mpc^{-3}$
is the mean mass density of the universe at redshift z.

We estimate the virial mass in the $\Lambda$CDM cosmology for the 63
clusters of the HIFLUGCS sample using Eq.~\ref{mteq}; the fit parameters 
($\beta$ and $r_c$ corrected for a $\Lambda$CDM cosmology) and the temperature $T$ 
are given in RB02. We have searched in the literature for $\beta$-fit parameters and $T$ of the clusters with GRHs (ref. in Tab.~\ref{beta_fit})
in order to estimate $M_v$ also for these clusters. Since some clusters of the 
HIFLUGCS sample are also in our sample, 
we note that in the majority of these cases the fits to the mass profile 
(and $T$) given in RB02 leads to a virial mass which is consistent 
at 1$\sigma$ level with the mass derived by making use of the parameters 
obtained from more recent observations in the literature (given in Tabs.~1, 2). 
The $L_x-M_v$ distribution of the combined sample is reported in Fig.~\ref{LMfig}.
The presence of a relatively large dispersion indicates 
the difficulty in estimating the virial masses of the single objects
and confirms the need of large samples in these studies. 
We note that the statistical distribution of clusters with GRHs 
is not different from that of the HIFLUGCS sample.
On the other hand, we note that clusters with known GRHs
span a range in mass comparable to the mass--dispersion
in the HIFLUGCS sample which is due to the different dynamical status
of clusters in the sample and to the uncertainties in the measurements.
This further strengthens the need of the approach followed in this 
Section, since a $L_x$ (or $P_{1.4}$)--$M_v$ fit based on GRHs
alone would be affected by large uncertainties.

In order to better sample the region of higher X-ray luminosities
and masses (typical of clusters with GRHs), we
compute the $L_x$--$M_v$ fit by combining the HIFLUGCS with the
radio--halo sample.
The fit has been performed using the form:

\begin{equation}  
\log\bigg[\frac{L_X}{10^{44}\,h_{70}^{-1}\,\frac{ergs}{s}}\bigg]=A_f+b_f\,
\log\Big(\frac{M_v}{3.16 \times 10^{14}\,h_{70}^{-1}\,M_{\odot}}\Big)
\label{LMeq}
\end{equation}

\noindent The best fit values of the parameters are: $A_f=-0.229\pm 0.051$
and $b_f=1.47\pm0.08$ ($b_f=1.41\pm0.10$ is obtained with HIFLUGCS sample only).

In order to derive the $P_{1.4\,GHz}-M_v$ correlation for GRHs,
we combine Eqs~\ref{LMeq} and \ref{LxLreq} and find :

\begin{eqnarray}
\lefteqn{\log\Big[\frac{P_{1.4}}{3.16\cdot10^{24}\,h_{70}^{-1}\,\frac{Watt}{Hz}}\Big]=(2.9\pm 0.4)
\log\Big[\frac{M_v}{10^{15}\,h_{70}^{-1}\,M_{\odot}}\Big] {} }
\nonumber\\
& & {} \;\;-(0.814\pm 0.147)
\label{LrMeq}
\end{eqnarray} 

Our $P_{1.4\,GHz}-M_v$ correlation
is slightly steeper than that obtained with 10 clusters
by Feretti (2003) ($P_{1.4\,GHz}\propto M^{2.3}$),
which, however, was derived in an EdS cosmology by considering the mass within 3 $h_{50}^{-1}$ Mpc 
from the cluster centers, and not the virial mass.

\section{Expected correlations and magnetic field constraints}

The main goal of this Section is to extract the values of the physical 
parameters to be used in the model calculations of Sec.4-6. The region 
of the physical parameters (in particular of B) is constrained by 
comparing the model expected and observed trends of the synchrotron 
power of GRHs with the mass (and temperature) of the parent clusters.
As already discussed (Sec.2) it is unlikely that the observed
correlations are driven by selection effects; however, it cannot be 
excluded that the detailed shape and scatter of these correlations 
might somewhat change with improved statistics, especially at low X-ray luminosities.

\subsection{Radio power--cluster mass correlation}

\begin{figure}
\resizebox{\hsize}{!}{\includegraphics{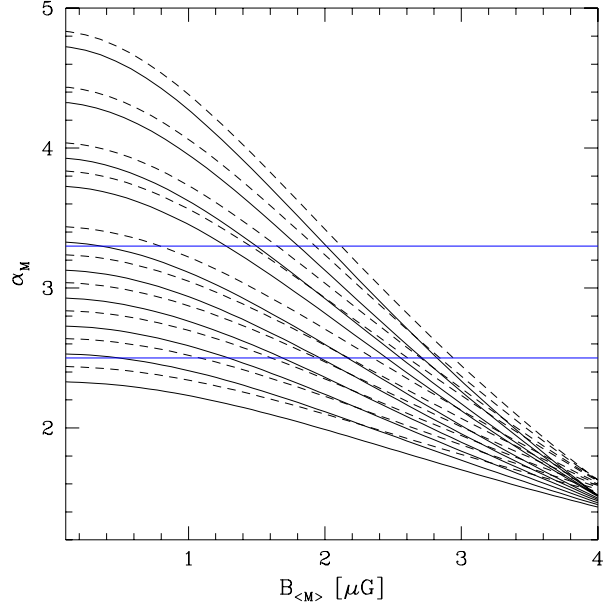}}
\caption[]{Expected slope of the $P_{1.4}-M_v$ correlation as a function of the magnetic 
field intensity in a cluster of mass $<M>=1.6\times\,10^{15}\,M_{\odot}$. 
The calculations are obtained for b=0.5,0.6,0.7,0.8,0.9,1,1.2,1.3,1.5 and 1.7 (from bottom to top); $M_1=1.1\times\,10^{15}\,M_{\odot}$ and $M_2=2.5\times\,10^{15}\,M_{\odot}$
are adopted. The continuous lines are for $\Gamma\simeq0.67$ and the dashed lines are for $\Gamma\simeq0.56$. The two
horizontal lines mark the 1 $\sigma$ value of the observed slope.}
\label{slopeM}
\end{figure}

\begin{figure}
\resizebox{\hsize}{!}{\includegraphics{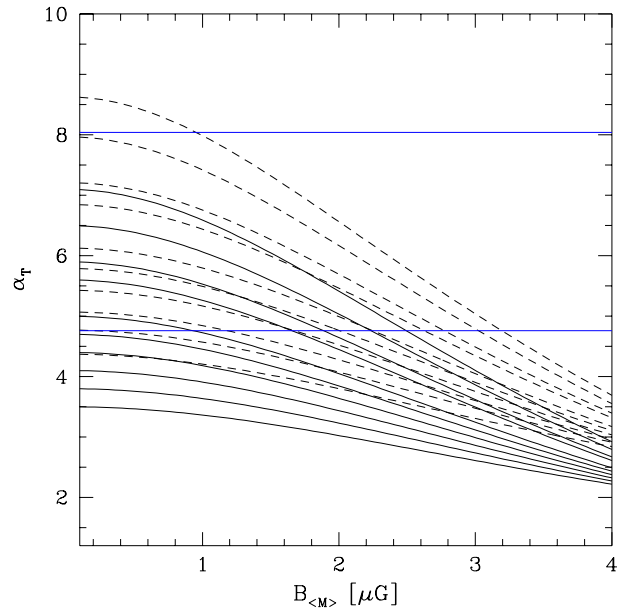}}
\caption[]{Expected slope of the $P_{1.4}-T$ correlation as a function of the magnetic 
field intensity in a cluster with temperature $<T>=8$ keV. The calculations are 
obtained for b=0.5,0.6,0.7,0.8,0.9,1,1.2,1.3,1.5 and 1.7 (from bottom to top);
 $T_1=6$ keV and $T_2=10$ keV are adopted. The continuous lines are for 
$\Gamma\simeq0.67$ and the dashed lines are for $\Gamma\simeq0.56$. The two
horizontal lines mark the 1 $\sigma$ value of the observed slope.}
\label{slopeT}
\end{figure}

Cassano \& Brunetti (2005) derived an expected trend between 
the bolometric radio power, $P_R$, 
and the virial cluster's mass and/or temperature.
In the case of the GRHs, the mergers which mainly
contribute to the injection of turbulence in the ICM are those with
$r_s\ge R_H$, $r_s$ being the stripping radius of the infalling sub--cluster 
(see Sec.~6 in CB05). It can be shown that, as a first approximation, 
the expected scaling $P_{R}-M_v$ is given by: 

\begin{equation}  
P_{R}\propto \frac{M_v^{2-\Gamma}\,B^2\,n_e}{(B^2+B_{cmb}^2)^2}
\label{PMpre}
\end{equation}

\noindent where $B$ is the rms magnetic field strength in the radio 
halo volume (particle pitch angle isotropitazion is assumed), 
$B_{cmb}=3.2 (1+z)^2 \mu G$ is the equivalent magnetic field 
strength of the CMB and $n_e$ is the number density of relativistic
 electrons in the volume of the GRH. The parameter $\Gamma$ 
is defined by $T\propto M^{\Gamma}$; we consider $\Gamma\simeq 2/3$ 
(virial scaling) and $\Gamma\simeq0.56$ (e.g.; Nevalainen et al. 2000).

\noindent In this paper we release the assumption adopted in CB05
of a magnetic field independent of cluster mass and 
assume that the rms field in the emitting volume scales as $B=B_{<M>}(M/<M>)^{b}$, 
with $b > 0$ and $B_{<M>}$ the value of the rms magnetic field associated to a cluster
with mass equal to the mean mass $<M>$ of the clusters sample.
A scaling of the magnetic field intensity with the cluster mass is 
indeed found in numerical cosmological MHD simulations (e.g. Dolag et al. 
2002, 2004). Dolag et al. (2002) found a scaling $B\propto T^2$ that would mean $B\propto M^{1.33}$ assuming the virial scaling or $B\propto M^{1.12}$ for $\Gamma\simeq0.56$.

\noindent
We assume that the number density of the relativistic electrons 
in galaxy clusters, $n_e$, does not depend on cluster mass. 
This is because there is no straightforward physical reason
to believe that this value should scale systematically with $M_v$,
and since only a relatively fast scaling of $n_e$ with mass
would significantly affect the radio power -- mass trend (Eq.~\ref{PMpre}).
It is indeed more likely that $n_e$ may change from cluster to cluster, 
but that the major effect would simply be to drive some scattering 
on the $P_R-M_v$ trend (Eq.~\ref{PMpre}).

\noindent
Given these assumptions Eq.~\ref{PMpre}
becomes:

\begin{equation}  
P_{R}\propto \frac{M_v^{2-\Gamma}\,B_{<M>}^2\cdot (M_v/<M>)^{2b}}
{(B_{<M>}^2\cdot(M_v/<M>)^{2\,b}+B_{cmb}^2)^2}
\label{PMpre2}
\end{equation}

\noindent    
which has two asymptotic behaviors: $P_R\propto M_v^{2-\Gamma+2b}$
for $B_{<M>}<< B_{cmb}$ and $P_R\propto M_v^{2-\Gamma-2b}$ for $B_{<M>}>>B_{cmb}$.

\noindent 
The observed correlations derived in Sect.~2 involve the
monochromatic radio power at 1.4 GHz. How this monochromatic radio
power can be scaled to $P_R$ depends on the spectrum of radio halos.
In the context of particle acceleration models (e.g., Brunetti et al. 2001, 
Ohno et al 2002, Kuo et al. 2003) the spectrum of radio halos is given by 
the superposition of spectra emitted from regions in the emitting volume
with different magnetic field strenghts. It is expected
to reach a peack at $\nu_b$ and then gradually drop
as a power-law which should further steepen at higher frequencies. The break
frequency can be expressed as a function of the cluster mass and of the rms field B
in the emitting volume(CB05):

\begin{equation}
\nu_b\propto M^{2-\Gamma}{{B\:\eta_t^2}\over{(B^2+B_{cmb}^{2})^{2}}}
\label{nub}
\end{equation}

\begin{figure}
\resizebox{\hsize}{!}{\includegraphics{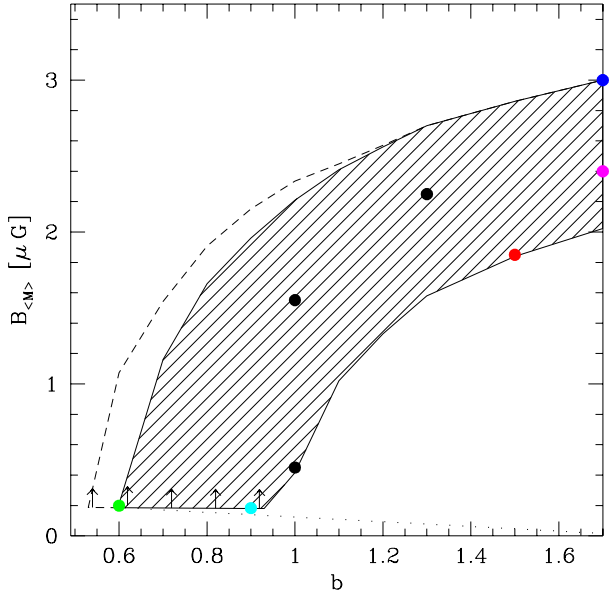}}
\caption[]{The region in the plane ($B_{<M>}$,b) allowed from the observed $P_{1.4}-M_v$ and $P_{1.4}-T$ 
correlations is reported as a shadowed area; $<M>=1.6\times\,10^{15}\,M_{\odot}$. The dashed line indicate the upper bound
of the allowed region obtained considering only the $P_{1.4}-M_v$ correlation. The coloured points 
indicate the relevant configurations of the parameters used in the statistical calculations in Sec.4-6 (Tab.~\ref{choose_value}). The vertical arrows
indicate the IC limits on $B$.}  
\label{regions}
\end{figure}

\noindent If we adopt a power-law spectrum extending from the frequency of the peak
to a few GHz, $P(\nu)\propto\nu^{-a}$, $P_R$ and the monochromatic radio power at a fixed frequency $\nu_o$ ($\nu_o\ge\nu_b$) scale as $P(\nu_o)/P_R\propto(\frac{\nu_b}{\nu_o})^{a-1}$.
This depends on the cluster mass (Eq.\ref{nub}):

\begin{equation}
\frac{P(\nu_o)}{P_R} \propto
{{ {M_v}^{(a-1)(2 - \Gamma +b)} }\over{
\left(B_{<M>}^2(M_v/<M>)^{2b}
+B_{cmb}^2\right)^{2(a-1)} }}
\label{setti}
\end{equation}

\noindent
thus in the case $B<<B_{cmb}$ one has $P(\nu_o)/P_R\propto(\frac{M}{<M>})^{(a-1)(2-\Gamma+b)}$,
while in the case $B>>B_{cmb}$ one has $P(\nu_o)/P_R\propto(\frac{M}{<M>})^{(a-1)(2-\Gamma-3b)}$, which means that for
$B<<B_{cmb}$ the $P(\nu_o)-M$ trend is steeper than the $P_R-M$, while the opposite
happens in the case $B>>B_{cmb}$ (the two scaling should be equal for continuity
for $B\sim B_{cmb}$). On the other hand, the trends of $P(\nu_o)/P_R$ with
the cluster mass in massive galaxy clusters is rather weak because the observed 
radio spectral index between 327--1400 MHz is $a \sim 1.2$ (e.g., Feretti 2003) and because B
in the most massive objects is probably close to $B_{cmb}$ (Sec.3.3, Fig.\ref{regions}; Govoni \& Feretti 2004).
Thus, in order to compare the model expectations with the observations,
we will safely assume the same scaling for monochromatic and total radio power.

In order to have a prompt comparison with observations we calculate 
the slope $\alpha_M$ of the $P_{1.4}-M$ correlation between two points as: 

\begin{equation}
\alpha_M=\frac{\log(P_1/P_2)}{\log(M_1/M_2)}
\label{alpha}
\end{equation}

\noindent Eq.\ref{alpha} can be compared with the observed slope
to constrain the value of the magnetic field and of the slope, $b$, of the scaling between $B$ and the cluster mass. The $M_1$
and $M_2$ values give the representative mass range  
spanned by the bulk of clusters with GRHs,
while $B_{cmb}$ should be calculated at the mean redshift
of our sample ($<z>\simeq 0.19$).
We point out that given $B_{<M>}$ and $b$, the values of B are fixed 
for all the values of the masses of the clusters in our sample.

In Fig.\ref{slopeM} we report the expected slope $\alpha_M$
(Eq.~\ref{alpha}) as a function of $B_{<M>}$. 
The different curves are obtained for different scaling-laws of the magnetic 
field with the cluster mass ($b=0.5$ to $1.7$, see caption).
Dashed lines refer to $\Gamma\simeq 0.56$ and solid lines to the virial case. 
The two blue horizontal lines (Fig.\ref{slopeM}) indicate the range of 
the observed slope ($\alpha_M=2.9\pm0.4$).

\noindent Fig.~\ref{slopeM} shows that there are values of $B_{<M>}$ and $b$ for 
which the expected slope is consistent with the observed one. 
As a first result we find that with increasing b the values of $B_{<M>}$ should increase in order to match the observations (for example, $b\sim 0.6$ requires $B_{<M>}\:\sim\:0.2\:-\:1.4\:\mu$G
while $b\sim1.7$ requires $B_{<M>}\:\sim\:2\:-\:3\:\mu$G). 
Finally, the asymptotic behavior of Eq.\ref{PMpre2}, combined with the observed correlation
(Eq.~\ref{LrMeq}) allows to immediately constrain b: for $B_{<M>}\,<< B_{cmb}$ one has
$0.58(0.53)<b<0.98(0.93)$ for the virial (non-virial) case, 
whereas in the case of $B_{<M>}\,>>B_{cmb}$ 
the model expectations cannot be reconciled with the observations.
 
\begin{table}
\begin{center}
\caption{Values of $\alpha_M$ and $\eta_t$ derived for relevant sets of b, $B_{<M>}[\mu G]$ parameters.}
\begin{tabular}{c|c|c|c|c}
\hline
\hline
 b   & $ B_{<M>}[\mu G]$  &$\alpha_M$ & $\eta_{min} $   &  $\eta_{max}$  \\
\hline
\hline
  1.7  &  3.0   & 2.5   & 0.19  &   0.2  \\
  1.7  &  2.2   & 3.22  & 0.17 &   0.2  \\
  1.5  &  1.9   & 3.3   & 0.15  &   0.2  \\ 
  1.3  &  2.25  & 2.84  & 0.15  &   0.2  \\
  1.0  &  1.55  & 2.96  & 0.16  &   0.21  \\
  1.0  &  0.45  & 3.3   & 0.29  &   0.33  \\
  0.9  &  0.18  & 3.23  & 0.39  &   0.44  \\
  0.6  &  0.2   & 2.63  & 0.38  &   0.44  \\
\hline
 \hline
\label{choose_value}
\end{tabular}
\end{center}
\end{table}

\begin{figure}
\resizebox{\hsize}{!}{\includegraphics{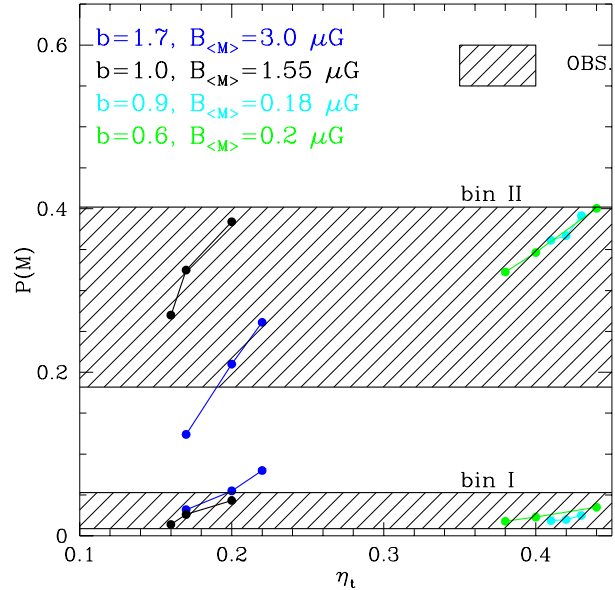}}
\caption[]{Probability to form GRHs at $ 0.05 \leq z \leq 0.15$ in the 
observed mass bin I: $0.95-1.9\times 10^{15} M_{\odot}$ and at 
$0.05 \leq z \leq 0.2$ in bin II: $1.9-3.8\times 10^{15} M_{\odot}$ as a function 
of $\eta_t$. The calculations are reported for the following representative 
cases: $b=1.7$, $B_{<M>}=3.0\mu$G (blue points); 
$b=1.0$, $B_{<M>}=1.55\mu$G (black points);  
$b=0.9$, $B_{<M>}=0.18\mu$G (cyan points) and 
$b=0.6$, $B_{<M>}=0.2\mu$G (green points). 
The bottom shadowed region marks the observed probability for GRHs 
in the mass bin I while the top shadowed region marks that in the mass bin II.
The values of the observed probabilities are
obtained by combining the results
from Giovannini et al. 1999, Giovannini \& Feretti 2000, and Feretti 2002.
The observed probabilities for the bin I are calculated up to $z \leq 0.15$ 
to minimize the effect due to the incompleteness of the X--ray and radio catalogs
used by these authors.}
\label{exa}
\end{figure}

\subsection{Radio power--cluster temperature correlation}

Since the temperature is related to the cluster mass, 
the radio power -- mass correlation also implies 
a correlation between synchrotron radio power and cluster temperature.
Thus, in order to maximize the observational constraints, an analysis similar 
to that of Sect.~3.1 can also be done for the 
radio power -- temperature correlation ($P_{R}-T$). 
Combining Eq.~\ref{PMpre2} with the $M-T$ scaling law
($T\propto M^{2/3}$ for the virial case and $T\propto M^{0.56}$) one has: 

\begin{equation}  
P_{R}\propto \frac{T^{\frac{2}{\Gamma}-1}\,B_{<M>}^2\,(T/<T>)^{2\,b_T}}
{(B_{<M>}^2\cdot(T/<T>)^{2\,b_T}+B_{cmb}^2)^2}
\label{PTpre}
\end{equation}

\noindent where $b_T=b/\Gamma$ with $\Gamma\simeq 2/3$ 
(virial case) or $\Gamma\simeq 0.56$ (non-virial case).
The asymptotic behaviors of Eq.~\ref{PTpre} are given by $P_R\propto T^{2/\Gamma-1+2b_T}$
($B_{<M>}\,<< B_{cmb}$) and $P_R\propto M_v^{2\Gamma-1-2b_T}$ 
($B_{<M>}\,>>B_{cmb}$). 

As in Sec.3.1, here we can adopt the same scaling with T for both
$P_{R}$ and $P_{1.4}$ and compare the values of the expected 
slope with those of the observed one.
We can calculate the slope $\alpha_T$ of the $P_{1.4}-T$ correlation between two
points as: 

\begin{equation}
\alpha_T=\frac{\log(P_1/P_2)}{\log(T_1/T_2)}
\label{alphaT}
\end{equation} 

\noindent where $T_1$ and $T_2$ define the interval 
of temperature of our sample, $<T>=8$ keV is the mean temperature,
and $B_{cmb}$ is evaluated at $<z>\simeq 0.19$.
In Fig.~\ref{slopeT} we report the slope $\alpha_T$ 
of the $P_{1.4}-T$ correlation as a function of the magnetic field 
strength in a average cluster, $B_{<M>}$.
The different curves are obtained for different scaling-laws of 
the cluster magnetic fields with mass (i.e., temperature)
(b=0.5 to 1.7). Dashed lines are for $\Gamma\simeq 0.65$ and continuous 
lines are for the virial case.

The horizontal blue lines mark the lower limit $\alpha_T\simeq 4.76$
and the upper limit $\alpha_T\simeq 8.05$ of the observed correlation.
Fig.~\ref{slopeT} shows that there is a range of values of the parameters
($B_{<M>},b$) for which the model is consistent with the observed slope.
The relevant point is that, similarly to the case of the $P_{1.4}-M$
correlations, also in this case values of $B_{<M>}\,>>B_{cmb}$
cannot be reconciled with observations: a clear upper boundary at $B<3\mu$G 
is obtained for $B_{<M>}$.

\begin{figure*}
\resizebox{\hsize}{!}{\includegraphics{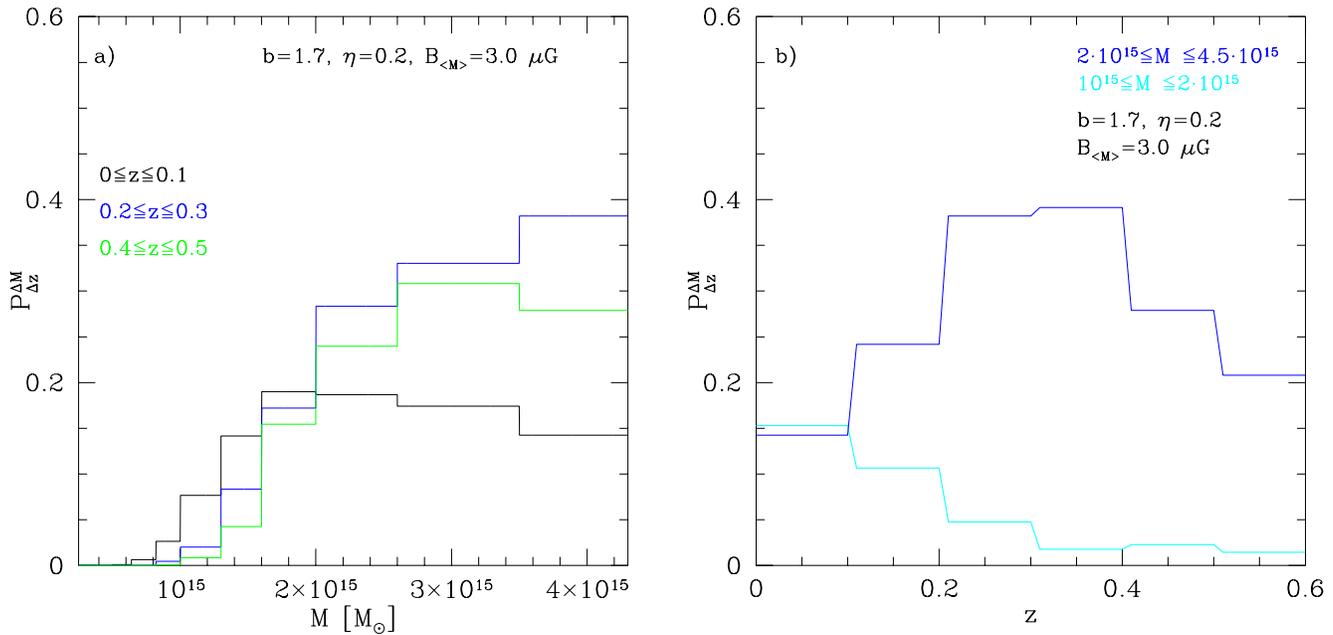}}
\caption[]{a) Occurrence of GRHs as a function of the cluster mass 
in three redshift bins:0-0.1 (black line),0.2-0.3 (blue line),
0.4-0.5 (green line). b) Occurrence of GRHs as a function of redshift 
in two mass bins: [1-2]$\times 10^{15} M_{\odot}$ (cyan line) and 
[2-4.5]$\times 10^{15} M_{\odot}$  (blue line).
The calculation have been performed assuming: b=1.7, $B_{<M>}=3.0 \mu$G, $\eta_t=0.2$ in both panels.}
\label{PMz_1p7}
\end{figure*} 

\subsection{Constraining the magnetic field}

We combine the results obtained from the observed correlations
(both $P_{1.4}-M_v$ and $P_{1.4}-T$) and the model expected trends to selects the allowed region of the 
($B_{<M>}$, $b$) parameters. We consider the slope of the $P_{1.4}-T$ correlation 
$\alpha_T\simeq 6.4\pm 1.64$ as derived for the extended sample of 24 galaxy clusters with 
giant and small radio halos. This is because what is important here is the allowed lower 
bound of the values of $\alpha_T$ which does not depend on the adopted sample (Sect.~2.2).

\noindent 
In Fig.\ref{regions} we report the region of the plane ($B_{<M>},b$) allowed by 
the observed slopes at 1$\sigma$ level. The lower bound of the ($B_{<M>}$,b) region is due to the
$P_{1.4}-M_v$ correlation 
while the upper bound is mostly due to the $P_{1.4}-T$ correlation which
is poorly constrained because of the very large statistical errors.
This bound is however also limited by the $P_{1.4}-M_v$ correlation (Fig.\ref{regions}, dashed line).

An additional limit on $B_{<M>}$, also reported in Fig.\ref{regions}
(vertical arrows), can be obtained from inverse Compton (IC) arguments. 
Indeed a lower bound to the magnetic field strength can be inferred 
in order to not overproduce, via IC scattering of the photons of the 
CMB radiation, the hard-X ray excess fluxes observed up to 
now in a few clusters (e.g., Rephaeli \& Gruber 2003, Fusco-Femiano et al 2003). 
In this case the value of the mean magnetic field intensity 
in the cluster volume can be estimated from the ratio between 
the hard-X ray and radio emission. 
The resulting value of the magnetic field 
should be considered as a 
lower limit because the IC emission may come from more external 
region with respect to the synchrotron emission 
(e.g., Brunetti et al. 2001, Kuo et al. 2003, Colafrancesco et al. 2005b)
and also because, in principle, 
additional mechanisms may contribute to the hard-X ray fluxes 
(e.g., Fusco-Femiano et al. 2003). 
One of the best studied cases is that of the Coma cluster for which 
an average magnetic field intensity of the order 
of $B_{IC}\simeq\,0.2\,\mu G$ was derived (Fusco-Femiano et al. 2004). 
As a first approximation we can use this value to obtain the lower bound 
of B for each cluster mass from the scaling $B=B_{<M>}(M/<M>)^{b}$.

The resulting ($B_{<M>}$,$b$) region spans a wide range of values 
of B and b. 
An inspection of Fig.\ref{regions} immediately identifies two allowed regimes:
a super-linear scaling ($b>1$) with relatively high values of B and a sub-linear scaling ($b<1$)
with lower values of B. 

All the calculations we will report in the following sections are carried 
out by assuming representative values of ($B_{<M>}$,$b$) inside the constrained region (Fig.~\ref{regions} coloured filled dots and Tab.\ref{choose_value}).

\section{Probability to form giant radio halos}

\subsection{Probability of radio halos and constraining $\eta_t$}

In this Section we derive the probability with cluster mass to find GRHs in the redshift range $z$=0--0.2.
The byproduct of the Section is to calibrate the model by requiring that
the expected fraction of cluster with GRHs is consistent with the observational
constraints. This allows to select a range of values of the parameter $\eta_t$,
which is the ratio between the 
energy injected in the form of magnetosonic waves and the  
$PdV$ work done by the infalling subclusters in passing through the 
most massive one. $\eta_t$ is a free parameter in our calculations
since the fraction of the energy which goes into the form of compressible
modes is likely to depend on the details of the driving force. 

\begin{figure*}
\resizebox{\hsize}{!}{\includegraphics{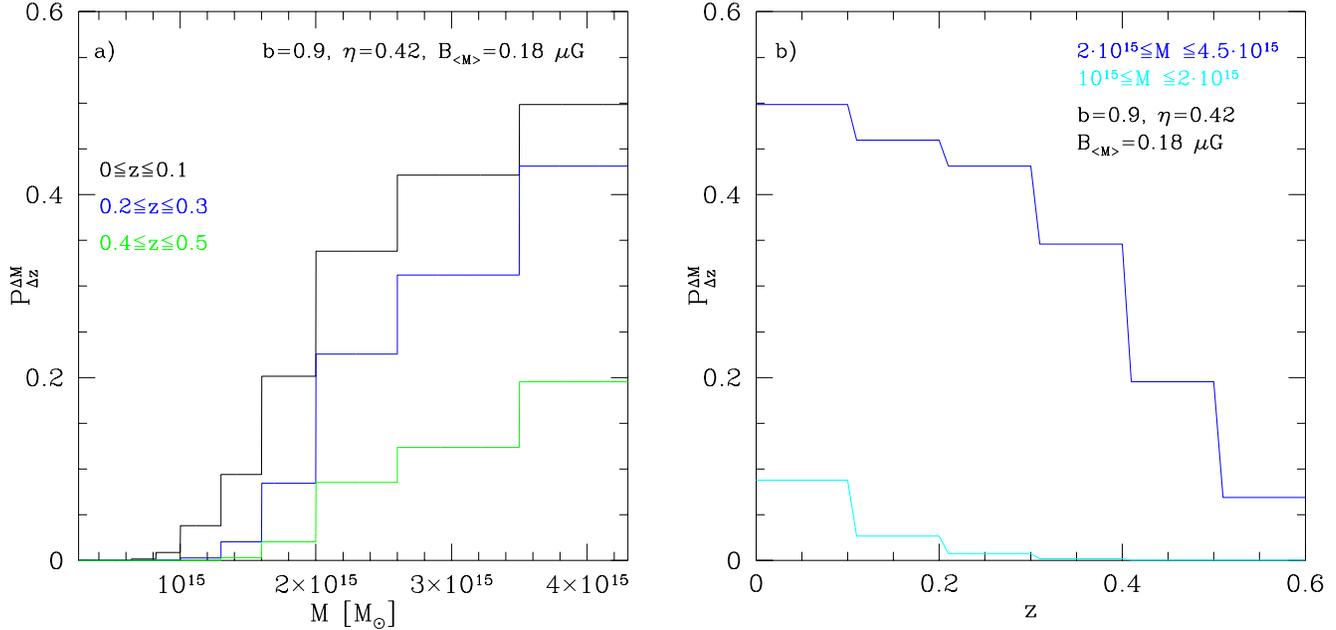}}
\caption[]{a) Occurrence of GRHs as a function of the cluster mass 
in three redshift bins: 0-0.1 (black line),0.2-0.3 (blue line),
0.4-0.5 (green line). b) Occurrence of GRHs as a function of redshift 
in two mass bins: [1-2]$\times 10^{15} M_{\odot}$ (cyan line) and
[2-4.5]$\times 10^{15} M_{\odot}$ (blue line).
The calculation have been performed assuming: b=0.9, $B_{<M>}=0.2 \mu$G, $\eta_t=0.42$ in both panels.}
\label{PMz_0p9}
\end{figure*}

\noindent In the conservative case of solenoidal forcing (and beta of plasma $>>$ 1) this fraction is expected to scale with $\mathcal{M}_s^2\,\mathcal{R}_e$ (with $\mathcal{M}_s<1$, the turbulent Mach number) for $\mathcal{M}_s^2\,\mathcal{R}_e\,<10$ and with a flatter slope for larger values
(Bertoglio et al. 2001). Assuming a Reynolds number (at the injection scale, \ie hundreds of Kpc)
in hot and magnetized galaxy clusters $\mathcal{R}_e\,\gtsim 10^3$ (see discussion in Lazarian 2006;
Brunetti 2006) and a turbulent energy
of the order of $\sim$ 20\% of the thermal energy (CB05), from Fig.~8 in Bertoglio et al. (2001) one finds a reference value $\eta_t\sim\,0.1$ which may be even larger in the case
of compressible driving.

\noindent Radio halos are identified with those objects in a synthetic cluster 
population with a synchrotron break frequency (Eq.\ref{nub}) $\nu_b\:\gtsim\: 200$ MHz
in a region of 1 $Mpc$ $h_{50}^{-1}$ size. 
In CB05 it was assumed that the magnetic field in 
the radio halo volume is independent from the cluster mass and it is 
$B\:\simeq\:0.5 \mu G$. Then $ \nu_b\: \propto \:M^{2-\Gamma}$
and consequently massive clusters are expected to be 
favourite in forming GRHs.
CB05 indeed showed that the expected fraction of clusters
with GRHs naturally shows an abrupt increase with
cluster mass, 
and that the observed fractions (20-30 \%
for $M > 2\times10^{15}\,M_{\odot}$ clusters, 2-5 \% for 
$M \sim 10^{15}\,M_{\odot}$ clusters and negligible for
less massive objects) can be well reconciled with the model
expectations by assuming $\eta_t \sim 0.24-0.34$.

In the present paper we assume that the rms magnetic field depends on the cluster mass 
and this should affect the synchrotron break frequency (Eq.~\ref{nub}) and the 
occurrence of GRHs with cluster mass.
 On the other hand, in Sect.~3.3 we have also shown that the comparison between
the expected and observed trends between radio power and
cluster mass (and temperature) helps in constraining the range of values
which can be assigned to the magnetic field in clusters.

\noindent Thus our calculations of the occurrence of GRHs ($z\le 0.2$) and the selection 
of the values of $\eta_t$ necessary to reproduce the observations should be performed 
within the dashed region in Fig.\ref{regions}.

To calculate the expected probabilities to form radio
halos we first run a large number,
${\cal{N}}$, of trees for different cluster masses at $z=0$, 
ranging from $\sim 5\times10^{14}M_{\odot}$ to 
$\sim 6\times10^{15} M_{\odot}$.
Then we choose different mass bins $\Delta M$
and redshift bins $\Delta z$ in which to perform our calculations.
Thus, for each mass $M$, we estimate the formation probability of GRHs
in the mass bin $\Delta M$ and in the redshift bin $\Delta z$ as (CB05):
 
\begin{equation}
f_M^{\Delta
M,\:\Delta z}=\frac{\sum_{j=1}^{{\cal{N}}}t_u^j}{\sum_{j=1}^{{{\cal{N}}}}
(t_u^j+t_d^j)}
\label{partialrate}
\end{equation}

\noindent where $t_u$ is the time in the redshift interval $\Delta z$
that the cluster spends in the mass bin $\Delta M$ with $\nu_b \geq\:200$MHz
and $t_d$ is the time that the same cluster spends in
$\Delta M$ with $\nu_b<200$ MHz.
The total probability of formation of GRHs in the mass
bin $\Delta M$
and in the redshift bin $\Delta z$ is obtained by combining
all the contributions (Eq.~\ref{partialrate}) weighted with the present
day mass function of clusters
given by the Press \& Schecther mass function.

To have a prompt comparison with present observational constraints, we calculate 
the probability to form GRHs at $z\,\ltsim \,0.2$ in the two observed mass bins:
bin I ($[0.95-1.9]\times 10^{15} M_{\odot}$) and
bin II ($[1.9-3.8]\times 10^{15} M_{\odot}$).

As an example, in Fig.~\ref{exa} we report these probabilities 
in both bin I and bin II as a function of $\eta_t$ 
for three representative cases which nicely sample the region in Fig.\ref{regions}: 
 $b=1.7$, $B_{<M>}=3.0\mu$G (blue points); $b=1.0$, $B_{<M>}=1.55\mu$G (black points); $b=0.9$, $B_{<M>}=0.18\mu$G (cyan points); $b=0.6$, $B_{<M>}=0.2\mu$G (green points).

The bottom shadowed region in Fig.~\ref{exa} marks the observed 
probability for GRHs in the mass bin I while the top shadowed region 
marks that in the mass bin II. Fig.~\ref{exa} shows that it is  
possible to find a range of values 
of the parameter $\eta_t$ for which the theoretical expectations are 
consistent with the observed statistics in both the mass bins. 
However we 
note that the requirement in terms of energy of the 
MS modes increases with decreasing the magnetic field:
it goes 
from $\eta_t \sim 0.15-0.2$ for intermediate--large values 
of $B$ up to $\eta_t \sim 0.5$ at the lower bound of
the allowed $B$ strengths.

The fact that the magnetic field depends on the cluster mass 
is reflected in the different behavior that the various 
selected configurations of parameters may have in Fig.~\ref{exa} 
in the two mass bins: one configuration of parameters
may be favoured in a mass bin with respect to another configuration
but disfavoured in the other mass bin.
This is related to the transition from IC dominance
($B < B_{cmb}$) to synchrotron dominance ($B > B_{cmb}$) that
occurs in going from the bin I to the more massive clusters of bin II.
In the case of IC dominance an increase of $B$ does not
significantly affect the particle energy losses, it causes
an increase of $\nu_b$ (Eq.\ref{nub}) and thus an increase of the 
probability to have GRHs.
On the other hand, in the case of synchrotron dominance 
the particle energy losses increase and consequently $\nu_b$ 
decreases (Eq.\ref{nub}) as well as the 
probability to form GRHs.

For this reason, given $\eta_t$, the ratio between the probability 
to form GRHs in the bin I and in the bin II is expected 
to decrease with increasing b, as larger values
of $b$ yield a more rapid 
increase of B with cluster mass (Fig.\ref{exa}).

In Tab.\ref{choose_value} we report the maximum and the minimum 
values of $\eta_t$ ($\eta_{t,max}$ 
and $\eta_{t,min}$) for which the model reproduces the observed probabilities 
(1 $\sigma$ limits) in both the mass bins. 
The results are given for the relevant ($B_{<M>},b$) configurations
reported in Fig.~\ref{regions}.
In agreement with the above discussions, one might notice that in the
case of IC dominance a larger magnetic field implies a smaller energetic 
request (smaller $\eta_{t,max}$).

\subsection{Probability of radio halos with $M_v$ and evolution with $z$}

In this Section we calculate the expected probability to
form GRHs with cluster mass 
without restricting ourselves to the mass bins considered by 
present observations (bin I and bin II in Fig.~\ref{exa})
and calculate the evolution of this probability with redshift.
In doing these calculations we use the values of $\eta_t$
as constrained in Tab.\ref{choose_value} within the region ($B_{<M>}$,b)
of Fig.\ref{regions}, and make the viable (and necessary)
assumption that the value of $\eta_t$ 
(i.e., efficiency of turbulence in going into MS modes)
is constant with redshift.

A detailed calculation of the acceleration efficiency and of 
the probability to have GRHs
requires detailed Montecarlo
calculations (see Sec.~6 of CB05) essentially because 
at each redshift the acceleration is driven by MS modes injected in
the ICM from the mergers that the cluster experienced in the last
few Gyr at that redshift.
However, to readily understand the model results reported in the following, we may use
the simplified formula Eq.~(\ref{nub}) which describes the approximate
trend of the break frequency with cluster mass. The scaling $B\propto M^b$ 
adopted in this paper implies that  
the synchrotron losses overcome the IC losses first in the more
massive objects.
Clusters of smaller mass in our synthetic populations have 
$B<<B_{cmb}$ and this implies (Eq.\ref{nub}) 
$\nu_b\propto M^{2-\Gamma+b}\,(1+z)^{-8}$ 
so that the probability to form GRHs in these clusters increases 
with the cluster mass ($2-\Gamma+b>0$ always) and decreases with redshift. 
In the case of more massive clusters the situation may be more complicated.
Indeed for these clusters there is a value of the mass, $M_{*}$, for which
the cluster magnetic field becomes equal to $B_{cmb}$. 
For $M\,>\,M_*(z)$ it is $\nu_b\propto M^{2-\Gamma-3b}$ (Eq.~\ref{nub})
and thus the probability to form GRHs would decrease as the mass becomes larger
(given the lower bound of the slope b as constrained in Fig.~\ref{regions}, 
it is $2-\Gamma-3b<0$). 
In these cases, at variance with the smaller clusters, the occurrence 
of GRHs with z is only driven by the cosmological evolution of the cluster-merger 
history (which drives the injection of turbulence) rather than by the dependence 
of the IC losses with z 
(at least up to a redshift for which $B\sim B_{cmb}(z)$). 
As a consequence, the general picture is that going from 
smaller to larger masses, the probability should reach a max value around
$M_*$ for which $B\sim B_{cmb}(z)$, and then it should start 
to smoothly decrease. 
The value of this mass increases with z and depends on the scaling 
law of B with M. It is:
 
\begin{equation}
M_*(z) \simeq\,<M> \left( {{3.2\, (1+z)^2}\over{B_{<M>}(\mu G)}}
\right)^{1/b}
\label{m*}
\end{equation}

In order to show in some detail this complex behavior in the following we 
analyze two relevant examples. 

\begin{figure}
\resizebox{\hsize}{!}{\includegraphics{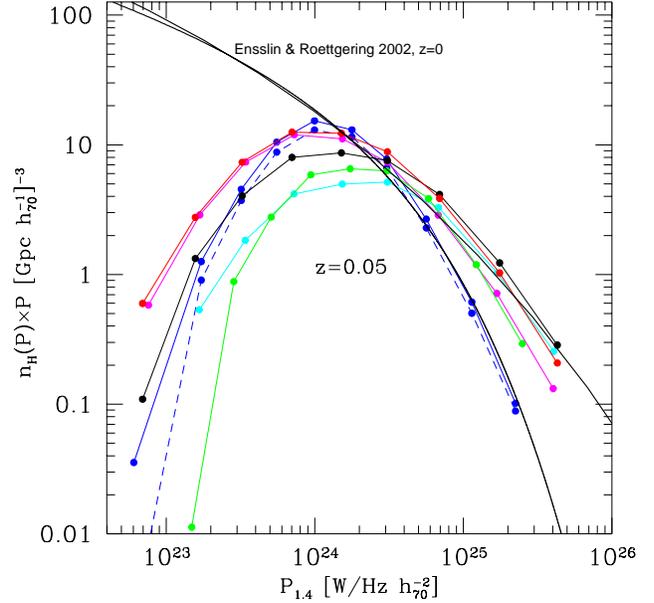}}
\caption[]{Expected RHLFs at $z\simeq0.05$ (coloured lines with dots)
obtained assuming:
b=1.7, $B_{<M>}=3.0\mu$G (blue lines: $\eta_t=0.2$ (solid line)
and $\eta_t=0.19$ (dashed line)); b=1.7, $B_{<M>}=2.2\mu$G and $\eta_t=0.2$
(magenta line); b=1.5, $B_{<M>}=1.9\mu$G and $\eta_t=0.2$ (red line);
b=0.9, $B_{<M>}=0.18\mu$G and $\eta_t=0.39$ (cyan line); 
b=0.6, $B_{<M>}=0.2\mu$G and $\eta_t=0.38$ (green line);
b=1.0, $B_{<M>}=0.45\mu$G and $\eta_t=0.33$ (black line).
For a comparison we report the range of Local RHLF obtained by 
E\&R02 (black solid thick lines).}
\label{RHLF_noi_ens}
\end{figure}

\subsubsection{An example with super-linear scaling: large B}
\begin{figure*}
\resizebox{\hsize}{!}{\includegraphics{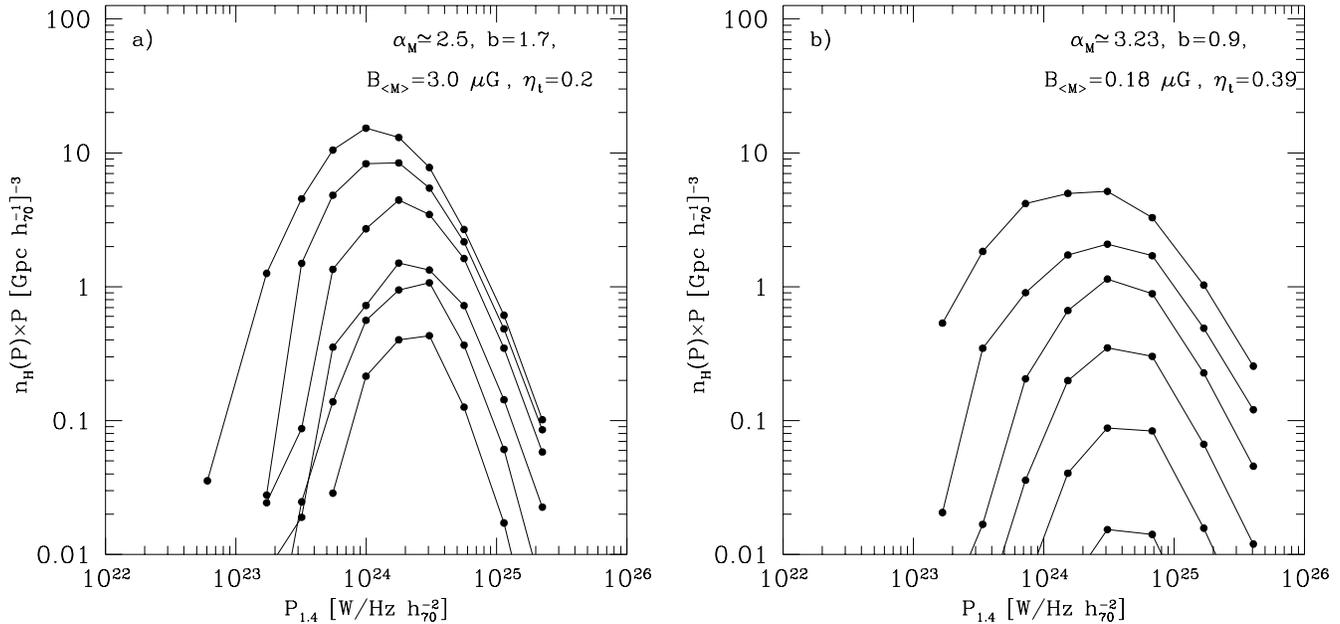}}
\caption[]{Evolution of RHLFs with redshift. The RHLFs are reported from
redshifts 0-0.1 to 0.5-0.6 (curves from top to bottom).
Calculations are developed for: Panel a) b=1.7, 
$B_{<M>}=3.0\,\mu$G, $\eta_t=0.2$, $\alpha_M\simeq 2.5$ and 
Panel b) b=0.9, $B_{<M>}=0.18\,\mu$G, $\eta_t=0.39$, $\alpha_M\simeq 3.23$.}
\label{RHLF_flat}
\end{figure*}
As a first example we focus on the case of a super-linear scaling. In Fig.~\ref{PMz_1p7}, 
we report the occurrence of GRHs as a function of the cluster mass in three redshift bins
(panel a)) and the occurrence of GRHs as a function of redshift 
in two mass bins (panel b)). 
These calculations have been performed using $b=1.7$
and $B_{<M>}=3 \mu$G which are allowed from the observed
correlations. We adopt $\eta_t=0.2$ which is in the 
corresponding range of values obtained in Sec.~5 (see Tab.~\ref{choose_value}) 
in order to reproduce the observed probability of 
GRHs at $z<0.2$. One finds that 
at lower redshifts ($z\,\ltsim\,0.1$) the probability to 
form GRHs increases with the mass of the clusters up to 
$M_*\:\sim\:2\,\times\,10^{15}\:M_{\odot}$, while for $M\:\gtsim M_*$
synchrotron losses become dominant and this causes
the decrease of the probability for $M\:\gtsim M_*$.
The mass at which $B\sim B_{cmb}(z)$ increases as $(1+z)^{2/b}$ and 
this causes the shift with z of the value of the cluster mass at which the maximum 
of the probability is reached.

\noindent Fig.\ref{PMz_1p7}b shows the occurrence of GRHs with z. In the 
higher mass bin ($2\cdot 10^{15}\le M \le 4.5\cdot 10^{15}$)
the occurrence increases up to $z\sim 0.4$ and than start 
to drop. In this very massive clusters the magnetic field is larger 
than $B_{cmb}(z)$ at any redshift and thus the synchrotron losses are 
always the dominant loss term. The behavior of the probability with z
in this case is essentially due to the fact that the bulk
of turbulence in these massive clusters is injected preferentially
between $z\sim 0.2-0.5$. A different behavior is observed in the lower 
mass bin ($10^{15}\le M \le 2\cdot 10^{15}$) where
 the occurrence of GRHs decreases with redshift. This is because 
 clusters with these lower masses have always $B<B_{cmb}(z)$.
 
\subsubsection{An example with sub-linear scaling: small B}

As a second example we focus on a sublinear scaling b.
In Fig.~\ref{PMz_0p9} we report the occurrence of GRHs as a function 
of the cluster mass in 
three redshift bins (panel a)) and the occurrence of GRHs as a function 
of redshift in two mass bins (panel b)). The calculations have been performed 
using $b=0.9$ and $B_{<M>}=0.2 \mu$G, which are allowed from the correlations, 
and adopting a corresponding $\eta_t=0.42$, which is within the 
range of values obtained in Sec.~5 (see Tab.~\ref{choose_value}) 
in order to reproduce the observed probability of formation of 
GRHs at redshift $z<0.2$.
In this case at any redshift the probability to form GRHs 
increases with the mass of the clusters. Indeed the magnetic field in these
 clusters is always $B<<B_{cmb}(z)$ (for all redshifts and masses) and 
 the IC losses are always the dominant loss term.
In addition, as expected, in both the considered mass bins 
the probability to form GRHs decreases as a function of redshift,
due to the increase of the IC losses (Fig.~\ref{PMz_0p9}, panel b)).

\section {Luminosity Functions of Giant Radio Halos}

In this Section we derive the expected luminosity functions of giant radio halos (RHLFs). 
Calculations for the RHLFs are carried out within the ($B_{<M>}$,b) region of Fig.~\ref{regions} by adopting the corresponding values of $\eta_t$ which allow 
to match the GRH occurrence at $z<0.2$.
First we use the probability $P_{\Delta z}^{\Delta M}$ to form GRHs with the cluster's 
mass to estimate the mass functions of GRHs ($dN_{H}(z)/dM dV$):
       
\begin{equation}
{dN_{H}(z)\over{dM\,dV}}=
{dN_{cl}(z)\over{dM\,dV}}\times P_{\Delta z}^{\Delta M}=n_{PS}\times
P_{\Delta z}^{\Delta M},
\label{RHMF}
\end{equation}

\begin{figure*}
{\hsize=5cm}{}{\includegraphics{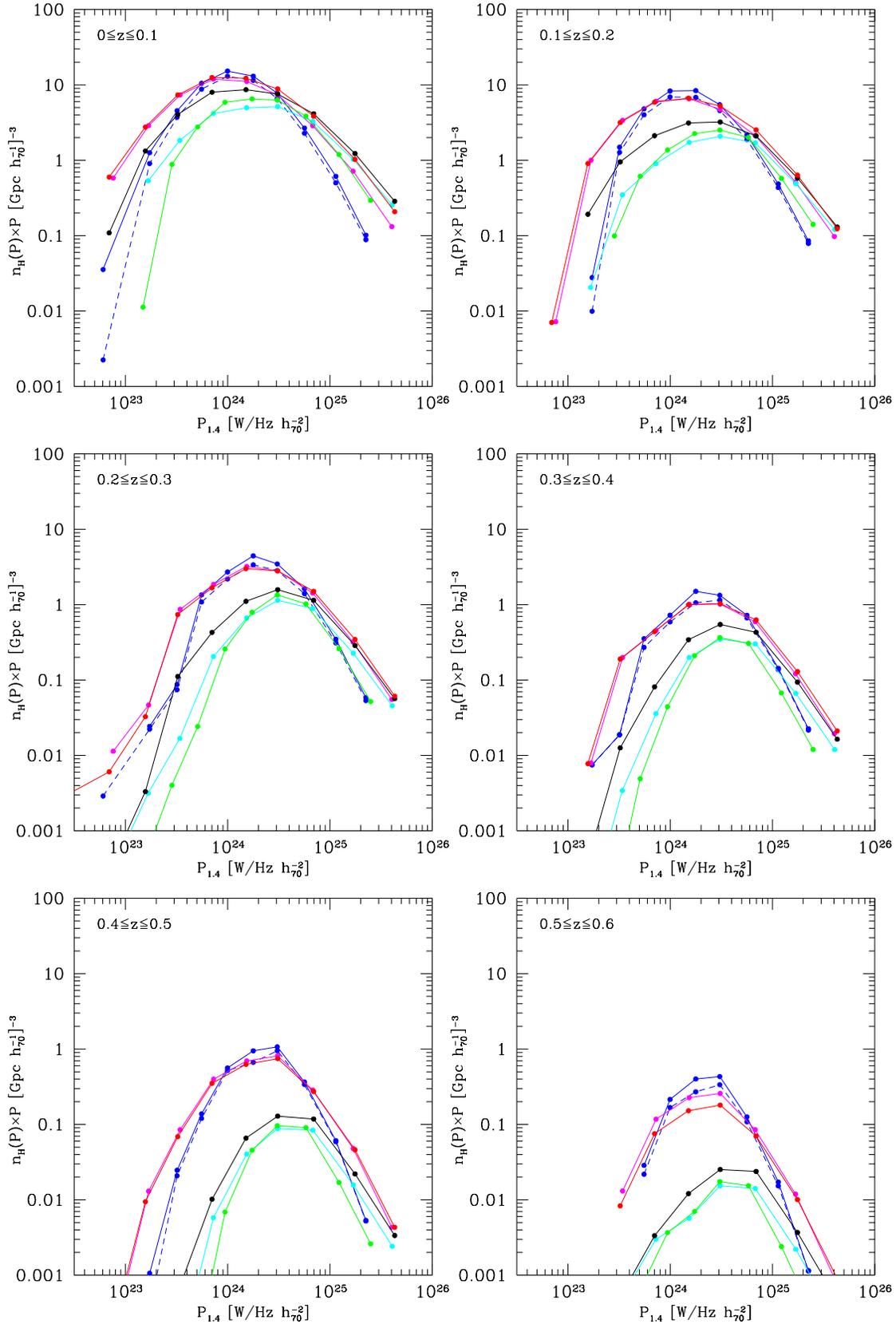}}
\caption[]{Expected RHLFs in 6 redshift bins (as reported in the panels). 
Calculations are performed by using the following values of the parameters :
b=1.7, $B_{<M>}=3.0\mu$G (blue lines: $\eta_t=0.2$ (solid lines)
and $\eta_t=0.19$ (dashed lines)); b=1.7, $B_{<M>}=2.2\mu$G and $\eta_t=0.2$
(magenta lines); b=1.5, $B_{<M>}=1.9\mu$G and $\eta_t=0.2$ (red lines); 
 b=0.9, $B_{<M>}=0.18\mu$G and $\eta_t=0.39$ (cyan lines);
b=0.6, $B_{<M>}=0.2\mu$G and $\eta_t=0.38$ (yellow lines); 
b=1.0, $B_{<M>}=0.45\mu$G and $\eta_t=0.33$ (black lines).}
\label{RHLF_flat_z}
\end{figure*}

\begin{figure}
\resizebox{\hsize}{!}{\includegraphics{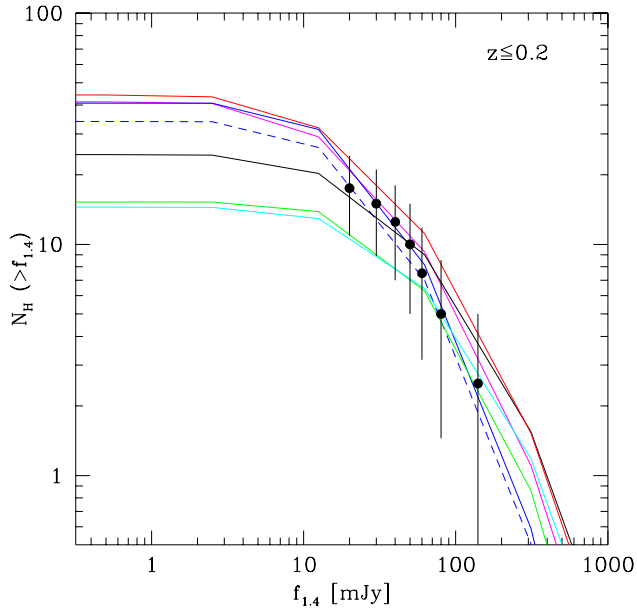}}
\caption[]{Number of expected GRHs above a given radio flux at 1.4 Ghz from a 
full sky coverage up to $z\le\,0.2$ (the colour code is that of Fig.\ref{RHLF_noi_ens}).
The black points are the data taken from Giovannini et al.(1999) and corrected 
for the incompleteness of their sky-coverage ($\sim\,2\,\pi$ sr).} 
\label{conteggi_z02}
\end{figure} 

\begin{figure}
\resizebox{\hsize}{!}{\includegraphics{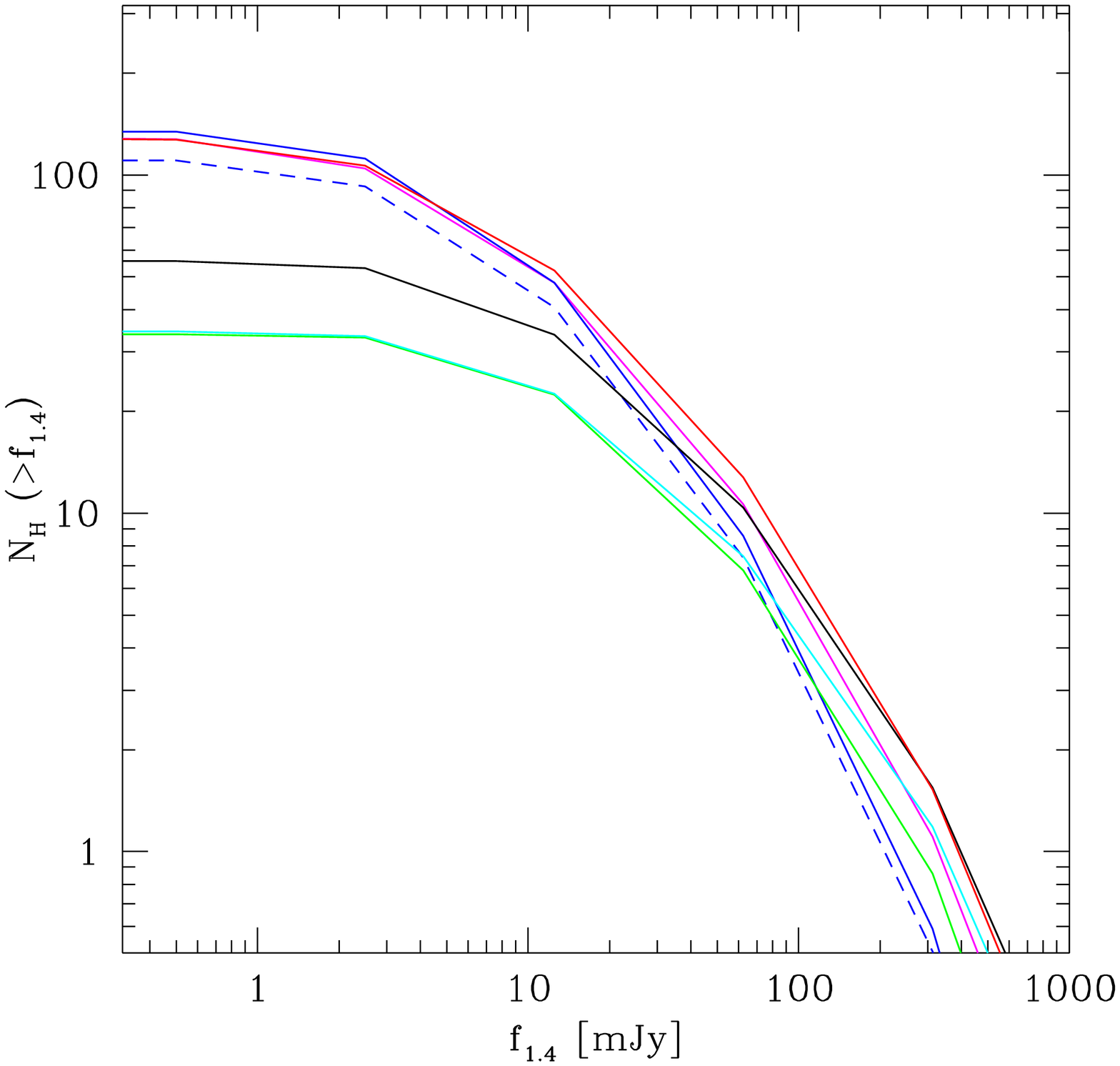}}
\caption[]{Number of expected GRHs from the whole universe 
above a given radio flux at 1.4 GHz. The colour code is the same of Fig.\ref{RHLF_noi_ens}.}
\label{conteggi_z06_f}
\end{figure} 

\noindent where $n_{PS}=n_{PS}(M,z)$ is the Press \& Schechter (1974) mass function
whose normalization depends essentially by $\sigma_8$ (present-day rms 
density fluctuation on a scale of $8 h^{-1}$ Mpc) and $\Omega_o$;
we use $\sigma_8=0.9$ in a $\Omega_o=0.3$ universe.
We stress that we use $n_{PS}$ since our model is based 
on the Press \& Schechter formalism. 

\noindent
The RHLF is thus given by:

\begin{equation}
{dN_{H}(z)\over{dV\,dP_{1.4}}}=
{dN_{H}(z)\over{dM\,dV}}\bigg/ {dP_{1.4}\over dM}.
\label{RHLF}
\end{equation}

\noindent
$dP_{1.4}/dM$ depends on the adopted ($B_{<M>}$, b) since 
each allowed configuration in Fig.~\ref{regions} selects
a value of the slope of $P_{1.4}-M_v$ (e.g., Tab.~3) which is consistent
 (at 1 $\sigma$) with the value of the observed slope 
 obtained with present observations ($\alpha_M=2.9 \pm 0.4$; see Sec.~3).
In particular from Fig.~\ref{slopeM} one has that, for a given $b$, 
larger values of the magnetic field select smaller values of the 
slope of the $P_{1.4}-M_v$ correlation (and viceversa).

In Fig.\ref{RHLF_noi_ens} we report the Local RHLFs 
(number of GRHs per comoving
$Gpc^3$ as a function of the radio power)
as expected from our calculations.
The most interesting feature in the RHLFs 
is the presence of a cut-off/flattening at low radio powers.
This flattening is a unique feature of particle acceleration models since it
marks the effect of the decrease of the efficiency of the particles acceleration
(in 1 Mpc $h_{50}^{-1}$ cube) in the case of the less massive galaxy clusters.
We stress that this result does not depend on the particular choice of the
parameters.

\noindent To highlight the result, in Fig.\ref{RHLF_noi_ens} we also compare
our RHLFs with the range of Local $(RHLFs)_{E\&R}$ (black solid lines)
reported by En\ss lin \& R$\ddot{o}$ttgering (2002).
These $(RHLFs)_{E\&R}$ are obtained by combining
the X-ray luminosity function of clusters with the radio-X-ray
correlation for GRHs and assuming that a costant fraction,
$f_{rh}=1/3$, of galaxy clusters have GRHs independently from the cluster mass
(see En\ss lin \& R$\ddot{o}$ttgering 2002).

\noindent The most important difference between the two expectations is
indeed that a low-radio power cut-off does not show up in the $(RHLFs)_{E\&R}$
in which indeed the bulk of GRHs is expected at very low
radio powers.
The agreement between the two Local RHLFs at higher synchrotron
powers is essentially because the derived occurrence of GRHs
in massive objects (Sect.~4.2) is in line with the fraction, $f_{rh}=1/3$, 
adopted by En\ss lin \& R$\ddot{o}$ttgering (2002).

In Fig.~\ref{RHLF_flat} we report the RHLFs 
expected by our calculations 
in different redshift bins. 
The calculations are performed by using 
two relevant sets of parameters (a super--linear and a sub--linear case as 
given in the caption 
of Fig.~\ref{RHLF_flat}) allowed from the observed correlations. 
With increasing redshift the RHLFs decrease due to the evolution of the clusters mass function 
with z and to the evolution of the probability to form GRHs with z. 

Fig.~\ref{RHLF_flat}, allows to readily appreciate the different behavior of the RHLFs
in the case of a super-linear scaling of B with M, $b=1.7$, 
(Fig.~\ref{RHLF_flat}, Panel a)) and of a sub-linear scaling,
$b=0.9$ (Fig.~\ref{RHLF_flat}, Panel b)): the evolution with redshift in the 
Panel b) (sub--linear case) is faster than that in the Panel a) (super--linear case).
This difference is driven by the probability to form GRHs as a function of redshift in the two cases: 
in the super--linear case the probability to form GRHs 
does not decrease rapidly with $z$, while a rapid decrease of such a
probability is obtained in the sub--linear case (see also Figs.~\ref{PMz_1p7}, ~\ref{PMz_0p9} Sect.~6).

In Fig.~\ref{RHLF_flat_z} we report the RHLFs obtained by our calculations by 
adopting the selected set of configurations given in Tab.~\ref{choose_value} (colour code is the same
of Fig.~\ref{regions}). 
The combination of these configurations define a bundle 
of expected RHLFs which determines the range of the possible RHLFs.

\begin{figure*}
\resizebox{\hsize}{!}{\includegraphics{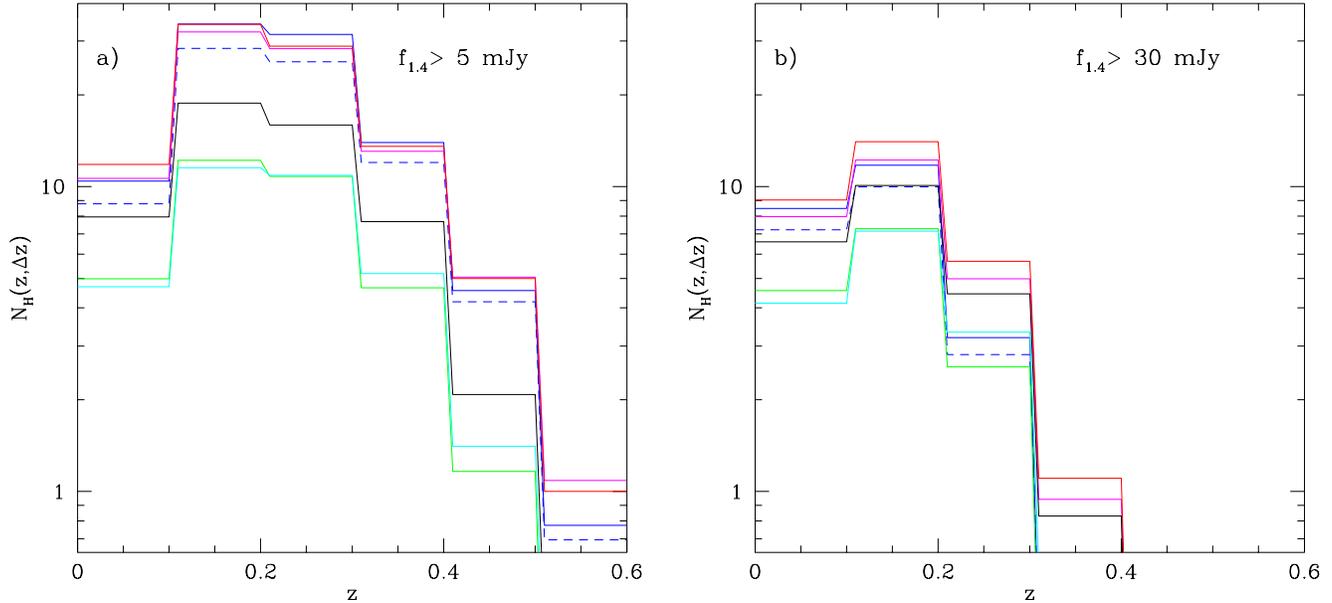}}
\caption[]{Expected total number of GRHs above a given radio flux 
in different redshift bins: panel a) above 5 mJy; panel b) above 30 mJy. 
In both panels the colour code is the same of Fig.\ref{RHLF_noi_ens}.}
\label{conteggi_z06}
\end{figure*} 

\noindent 
All the calculations are performed for the corresponding range of values of $\eta_t$ 
which allow to be consistent with the observed probability to form radio 
halos at $z\,\ltsim\,0.2$. One finds that with increasing 
redshift 
the bundle of the RHLFs broadens along the $n_{H}(P)\,\times\,P$ axis.
This is again due to the different evolutions of the probability to form GRHs 
with $z$ of the super--linear and sub--linear cases.

\section{Number counts of giant radio halos}

In this Section we derive the expected number counts of
giant radio halos (RHNCs). This will allow us to perform a first comparison 
between the model expectations and the counts of GRHs which can be
derived from present observations, but also to derive expectations for future
observations. 
As for the case of the RHLFs, in calculating the RHNCs we adopt the configurations of parameters
which allow to reproduce the observed probabilities of GRHs at $z<0.2$.
However, 
we point out that the fact that our expectations are consistent 
with the observed probability to form GRHs at $z\ltsim\,0.2$ 
does not imply that they should also be consistent with the observed flux distribution
of GRHs in the same redshift interval.

Given the RHLFs ($dN_H(z)/dP_{1.4}dV$) the number of GRHs 
with $f> f_{1.4}$ is given by:

\begin{equation}
N_{H}(>f_{1.4})=\int_{z=0}^{z}dz' ({{dV}\over{dz'}})
\int_{P_{1.4}(f_{1.4}^{*},z')}{{dN_H(P_{1.4},z')}\over{dP_{1.4}\,dV}}dP_{1.4}  
\label{RHNC}
\end{equation}

\noindent where $dV/dz$ is the comoving volume element in the $\Lambda$CDM cosmology
\noindent (e.g., Carroll, Press and Turner 1992); the radio flux and the radio
power are related by $P_{1.4}=4\pi\,d_L^2\,f_{1.4}$ with $d_L$ the luminosity distance
(where we neglect 
the K-correction since the slope of the spectrum of radio halos is close 
to unity).

As a first step, we use Eq.~\ref{RHNC} to calculate the number 
of expected GRHs above a given radio flux at 1.4 Ghz from a 
full sky coverage up to $z\ltsim0.2$ and compare the results with 
number counts derived by making use of the present 
day observations (Fig.~\ref{conteggi_z02}, 
the colour code is that of Fig.\ref{RHLF_noi_ens}).
Calculations in Fig.~\ref{conteggi_z02} are obtained by 
using the full bundle of RHLFs obtained in the previous Section
(Fig.~\ref{RHLF_flat_z}).
The black points are obtained by making use
of the radio data from the analysis
of the radio survey NVSS by Giovannini et al.(1999); normalization of
counts is scaled to correct for the 
incompleteness due to the sky-coverage in Giovannini et al.
($\sim\,2\,\pi$ sr).
The NVSS has a 1$\sigma$ level at 1.4 GHz equal to 
0.45 mJy/beam (beam=45$\times$45 arcsec, 
Condon et al. 1998). 
By adopting a typical size of GRH of the order of 1 Mpc, 
the surface brightness of the objects which populate
the peak of the RHLFs ($\sim 10^{24}$ W/Hz)
at z$\sim$0.15 is expected to fall below the 2$\sigma$ 
limit of the NVSS.
These GRHs have a flux of about 20 mJy, thus 
below this flux the NVSS becomes poorly efficient in catching
the bulk of GRHs in the redshift bin z=0--0.2 and a fair
comparison with observations is not possible.
For larger fluxes we find that the expected number counts are 
in excellent agreement with the counts obtained from the 
observations.
We note that assuming a superlinear scaling of $B$ with cluster
mass, up to 30-40 GRHs at $z<0.2$ are expected to be discovered with
future deeper radio surveys. On the other hand,  
the number of these GRHs in the case of a sublinear scaling
should only be a factor of $\sim 2$ larger than that of 
presently known halos.

As a second step, we calculate (Fig.\ref{conteggi_z06_f}) the whole sky number 
of GRHs expected up $z=0.7$ (the probability to form GRHs at $z\,>\,0.7$ is negligible). 
We note that the number counts of GRHs increases down to
a radio flux 

\begin{figure*}
\resizebox{\hsize}{!}{\includegraphics{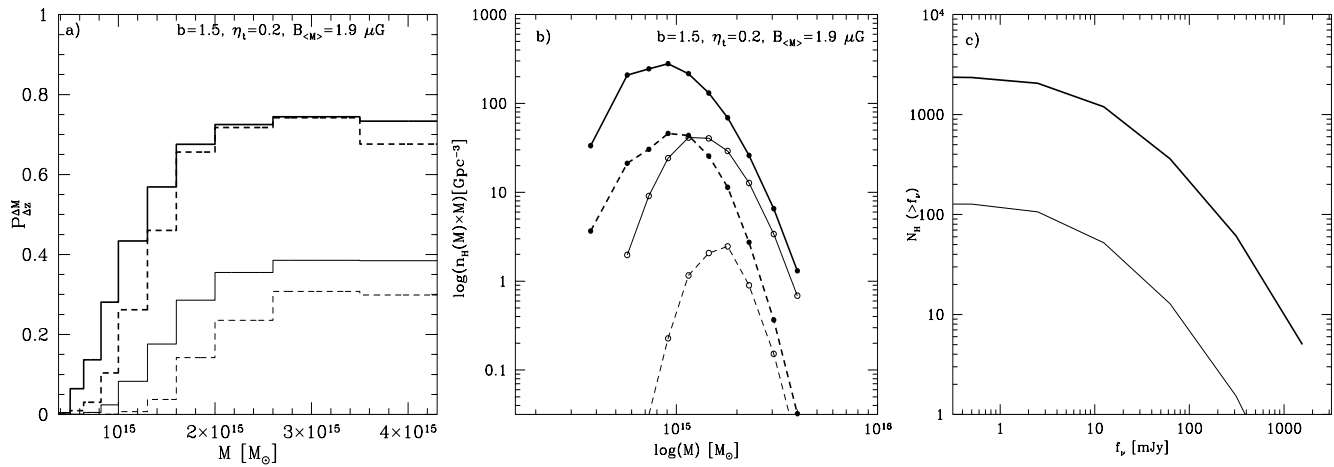}}
\caption[]{{\bf a)} The occurrences of GRHs as a function of the cluster mass 
in the redshift bins 0-0.1 (solid lines) and 0.4-0.5 (dashed lines) are reported
for 150 MHz (thick lines) and for 1.4 GHz (thin lines).
{\bf b)} Mass functions of GRHs in the redshift bins 0-0.1 (solid lines) and 
0.4-0.5 (dashed lines) are reported for 150 MHz (thick lines) and for 1.4 GHz (thin lines).
{\bf c)} Comparison between the expected RHNCs above a given radio flux at 1.4 Ghz 
(thin lines) and at 150 MHz (thick lines) from a full sky coverage up to 
$z\le\,0.6$. \\
All the calculations have been performed assuming: 
b=1.5, $B_{<M>}=1.9 \mu$G and $\eta_t=0.2$.
}
\label{lofar}
\end{figure*} 

\noindent of $f_{1.4}\sim\,2-3$ mJy and then flattens due to the 
strong (negative) evolution of the RHLFs (Fig.~\ref{RHLF_flat_z}). 
We note that the expected total number of GRHs above 1 mJy at 1.4 GHz 
is of the order of $\sim\,100$ depending on the scaling 
of the magnetic field with cluster mass.

Finally we calculate the expected number counts of GRHs
above a given radio flux in different redshift bins. This allows us to 
catch the redshift at which the bulk of GRHs is expected.
In Fig.~\ref{conteggi_z06} we report the RHNCs integrated 
above 5 mJy (Panel a)) and above 30 mJy (Panel b)). We note that the bulk 
of GRHs is expected in the redshift interval $0.1-0.3$
and this does not strongly depend on the flux limit. 
We note that the relatively high value of such redshift range 
is also due to the presence of the 
low radio power cut-off in the RHLFs which suppresses the expected number of 
low power GRHs. On the other hand, at radio fluxes $>$ 30 mJy the 
contribution from higher redshift decreases since the requested radio luminosities
at these redshift correspond to masses of the parent clusters which are above
the high--mass cut-off of the cluster mass function.

\section{Towards low radio frequencies: model expectations at 150 MHz}

Due to their steep radio-spectra, GRHs are ideal
targets for upcoming low-frequency radio telescopes, 
such as LOFAR and LWA. In this section we present 
calculations of the statistics of GRHs at 150 MHz 
derived from the electron reacceleration model.

For simplicity, we present these results only for 
one set of the parameters in the plane ($B_{<M>},b$) (Fig.\ref{regions}): 
a super-linear case (b=1.5, $B_{<M>}=1.9 \mu$G) (see Sec.3). 

First, we calculate the probability to have GRHs at $\sim 150$ MHz
as a function of the cluster's mass following the procedure outlined
in Sec.4.1 and requiring a break frequency $\nu_b\gtsim 20\,$ MHz to account for the observation frequency. In Fig.\ref{lofar}a we report the probability to have  
GRHs as a function of virial mass in two redshift bins at 1.4 GHz 
(thin lines) and at 150 MHz (thick lines). As expected, the probability  
at 150 MHz is substantially larger than that calculated at 1.4 GHz,
particularly for higher redshifts and for low massive clusters.

One of the main findings of our work is the presence of a cut-off 
in the RHLFs at low radio powers (see Sec.5), which reflects the drop of 
the probability to form GRHs as the cluster's mass decreases.
In Fig.\ref{lofar}b we plot the mass functions of radio 
halos (RHMFs) at 1.4 GHz and at 150 MHz in two redshift bins
(see caption of Fig.\ref{lofar}). We note that the number density
of GRHs is increased by only a factor $\sim 2$ 
for $M> 2\cdot 10^{15}\,M_{\odot}$, but by more than one 
order of magnitude for $M\le\,10^{15}\,M_{\odot}$.
The most interesting feature is again the presence 
of a low mass cut-off in the RHMFs at 150 MHz, which however 
is shifted by a factor $\sim 2$ towards smaller masses with 
respect to the case at 1.4 GHz.
This is related to the fact that a smaller energy density in the form
of turbulence is sufficient to boost GRHs at lower frequencies,
and this allows the formation of GRHs also in slightly smaller clusters,
which indeed are expected to be less turbulent (CB05; see also Vazza et al. 2006). 

Finally, in order to obtain estimates for the RHLFs and RHNCs 
at 150 MHz, we tentatively assume the same $P_{R}-M$ scaling found
at 1.4 GHz, scaled at 150 MHz with an average spectral index 
$\alpha_{\nu}\sim 1.2$, and follow the approach outlined in Secs.~5 and 6.
In Fig.\ref{lofar}c we report the expected integral number counts 
of radio halos from a full sky coverage above a given radio flux at 1.4 GHz (thin lines) 
and at 150 MHz (thick lines) up to a redshift $z\sim\,0.6$. 
The expected number of GRHs at 150 MHz are a factor of $\sim 10$
larger than the number expected at 1.4 GHz, with the bulk of GRHs
at fluxes $\ge$ few mJy. In the near future LOFAR will be able to 
detect diffuse emission on Mpc scale at 150 MHz down to these fluxes and this would be 
sufficient to catch the bulk of these GRHs (a more detailed study will
be presented in a forthcoming paper).

\section{Summary and Discussion}

The observed correlations between radio and X-ray properties of galaxy clusters
provide useful tools to constraining the physical parameters that are relevant
to the reacceleration models for the onset of giant radio halos (GRHs). Our analysis 
is based on the calculations of Cassano \& Brunetti (2005; CB05), which assumes that a 
seed population of relativistic electrons reaccelerated by magnetosonic 
(MS) waves is released in the ICM by relatively recent merger events. To this 
end we have collected from the literature a sample of 17 GRH clusters for all of which,
but one (A2254), both radio and X-ray homogeneous data are available, as summarized 
in Tab.1 \& 2. Based on the relationships derived in CB05 paper, we have been able to constrain the (likely) dependence of the average magnetic field intensity (B) on the cluster mass, under the assumption that B can be parameterized as $B =B_{<M>} (M/<M>)^b$ (with $B_{<M>}$ the average field intensity of a cluster of mean mass 
$<M>=1.6\times\,10^{15}\,M_{\odot}$ and b positive). This is an important achievement because both the emitted synchrotron spectrum and losses 
depend critically on the field intensity. Following CB05 approach, the merger
events are obtained in the statistical scenario provided by the extended 
Press \& Schechter formalism that describes the hierarchical formation of galaxy clusters. The main results of our study can be summarized as follows:

\begin{itemize}

\item[$\bullet$] {\it Observed correlations} 
\\

In Sect.2 we derive the correlations between the radio power at 1.4 GHz
($P_{1.4}$) and the X-ray luminosity (0.1-2.4 keV), ICM temperature and cluster mass. 
Most important for the purpose of the present investigation is the $P_{1.4}-M_v$
correlation which has been derived by combining the $L_X-M_v$
correlation obtained for a large statistical sample of galaxy clusters
(the HIFLUGCS sample plus our sample) with the $P_{1.4}-L_X$ correlation derived 
for our sample of GRHs. This procedure allows us to avoid the 
well known uncertainties and limits which 
are introduced in measuring the masses of small samples
of galaxy clusters, especially in the case of merging systems.
We find a value of the slope $\alpha_M=2.9 \pm 0.4$ ($P_{1.4} \propto M_v^{\alpha_M}$).
A steep correlation of the synchrotron luminosity with the ICM temperature
is also found, although with a large statistical
error in the determination of the slope : $\alpha_T=6.4 \pm 1.6$
($P_{1.4} \propto T^{\alpha_T}$).

In Sec.2 we have also shown that at least in the case of high X-ray
luminosity clusters ($L_X\gtsim\,5\cdot 10^{44}$ erg/s) the above trends are
unlikely driven by selection effects in the present observations.\\

\item[$\bullet$] {\it Constraining the magnetic field dependence on the mass}
\\

A correlation between the radio power and
the cluster virial mass is naturally expected in the 
framework of electron acceleration
models. This relationships, discussed in Sec. 3.1 (Eq.\ref{setti}),
can reproduce the observed correlation for viable values of the physical 
parameters. For instance, in the case
$B<<B_{cmb}$, it is $P(\nu_o)
\propto {M_v}^{a(2 - \Gamma +b)+b}$ 
and the exponent agrees with the observed one ($\alpha_M\sim 3$)
by adopting a typical slope of the radio spectrum $a=1-1.2$ 
and a sub--linear scaling $b\sim 0.6-0.8$.

A systematic comparison of the expected correlations between the 
radio power and the cluster mass with the observed one (Sect.3.1 \& 2) allows 
the definition of a permitted region of the parameters' space ($B_{<M>}$,b), 
where a lower bound $B_{<M>}=0.2\, \mu$G is obtained in order not to overproduce 
via the IC scattering of the CMB photons the hard X-ray fluxes observed 
in the direction of a few GRHs (Sect. 3.3 and Fig.6). It is found a lower bound 
at $b\sim 0.5-0.6$ and that a relatively narrow range of $B_{<M>}$ values is allowed for a fixed b. 
The boundaries of the allowed region, aside from the lower bound of
$B_{<M>}$, are essentially sensitive to the limits from the
$P_{1.4}-M_v$ correlation. 

\noindent A super--linear scaling of $B$ with mass, as expected by 
MHD simulations (Dolag et al.~2004)
falls within the allowed region.

\noindent
The values of the average magnetic field intensity in the superlinear
 case are close (slightly smaller) to 
those obtained from the Faraday rotation measurements (e.g., Govoni \& Feretti 2004), which, however,
generally sample regions which are even more internally placed than
those spanned by GRHs. 

Future observations will allow to better constrain the radio-X ray 
correlations and thus to better define the region of the model parameters.\\

\item[$\bullet$] {\it Probability to form GRHs}
\\

In Sect.4 we report on extensive calculations aimed at constraining $\eta_t$, 
the fraction of the available energy in MS waves, which is required to
match the observed occurrence of GRHs at redshifts $z\le 0.2$ (Fig.7). By adopting a 
representative sampling of the allowed ($B_{<M>}$,$b$) parameter space (Fig.6) we find 
$0.15 \le \eta_t\le 0.44$: the larger values are obtained for $B_{<M>}$ 
approaching the lower bound of the allowed region, because of the larger 
acceleration efficiency necessary to boost electrons 
at higher energies to obtain a fixed fraction of clusters with 
GRHs.

With an appropriate $\eta_t$ value for each set of ($B_{<M>}$,b) parameters we 
can calculate the probability of occurrence of GRHs at larger redshifts 
for which observational data are not available. 
This probability depends on the merging history of clusters  
and on the relative importance of the synchrotron and IC losses,
and shows a somewhat complicated behavior with cluster
mass and redshift. The maximum value of this probability at a given redshift is found 
for a cluster mass $M_*$ (Eq.\ref{m*}) which mark the transition 
between the Compton and the synchrotron dominated phases.

\noindent
In the case of sublinear scaling of the magnetic field
with cluster mass (b$\sim$0.6--0.9)
the allowed values of the strength of the 
magnetic field are relatively small (Fig.~\ref{regions}),
the value of $M_*$ is large
and the IC losses are always dominant for the
mass range of clusters with known GRHs.
As a consequence the probability to have GRHs
increases with cluster mass and decreases with redshift
(Fig~\ref{PMz_0p9}).
On the other hand superlinear scalings (b$\sim$1.2--1.7)
imply allowed values of $B_{<M>}$ relatively large (Fig.~\ref{regions}),
and even larger values of the magnetic field
for the most massive objects. 
In this case the value $M_*$ falls within the range of masses
spanned by GRH clusters: 
the predicted fraction of clusters with GRHs increases
with mass, then reaches a 
maximum value at about $M_v \sim M_*$, and finally
falls down for larger masses (Fig~\ref{PMz_1p7}).
At variance with the case of sublinear scaling, in this case
the fraction of the most massive objects with GRHs
is expected to slightly increase
with redshift, at least up to z=0.2--0.4 (Fig~\ref{PMz_1p7})
where the bulk of
turbulence is injected in a $\Lambda$CDM model (CB05).\\

\item[$\bullet$]{\it Luminosity functions (RHLFs)}
\\

In Sect.5 we report the results of extensive calculations following a fair sampling
of the ($B_{<M>}$,b) allowed region as summarized in Tab. 3; this essentially allows 
a full coverage of all possible RHLFs given the present correlations at 1$\sigma$. We find that, although the large uncertainties in the ($B_{<M>}$,b) region, the predicted local RHLFs are confined to a rather narrow bundle, the most characteristic common feature being the presence of a flattening/cut-off at radio powers below about $10^{24}$ W/Hz at 1.4 GHz (Fig.\ref{RHLF_noi_ens}). 
The fraction of GRHs with 1.4 GHz luminosity below $\sim 5 \times
10^{22}$W Hz$^{-1}$ h$_{70}^{-2}$, a factor of $\sim 5$ smaller
than the luminosity of the less powerful GRH (A2256, z=0.0581) known so far,
is negligible. This characteristic shape of the RHLFs,  
obtained in our paper for the first time, represents a unique prediction of particle
acceleration models, and does not depend on the adopted physical details for the particle
acceleration mechanism. This is due to the decrease of the efficiency of particle acceleration in the case of less massive clusters which is related to three 
major reasons (see CB05):
\begin{itemize}
\item[{\it i})] smaller clusters are less turbulent than larger ones since
the turbulent energy is expected to scale with the thermal one (CB05; see also
Vazza et al. 2006);
\item[{\it ii})] turbulence is typically injected in large Mpc regions in more
massive clusters and thus these are favoured for the formation of GRHs (CB05);
\item[{\it iii})] since in the  present paper we found $B\propto M^b$ with $b\gtsim\,0.5$,
higher energy electrons should be accelerated in smaller clusters to emit synchrotron
radiation at a given frequency.
\end{itemize}

\noindent Deep radio survey with future radio telescopes (LOFAR, LWA, SKA) 
are required to test the presence of this cut-off/flattening 
in the luminosity function of the GRHs.

The predicted evolution of the RHLFs with redshift is 
illustrated in Fig.~\ref{RHLF_flat_z}: the comoving number density of GRHs 
decreases with redshift due to the evolutions of the cluster mass
function and of the probability to form GRHs.
The decrease with redshift of the RHLFs calculated by adopting sublinear
scaling of the magnetic field with cluster mass is faster than
that in the superlinear scaling causing a spread in the RHLFs bundle
with z.\\

\item[$\bullet$]{\it Number counts (RHNCs)}
\\

In Sec.7 we have derived the integral number counts of GRHs at 1.4 GHz.
We find that the number counts predicted for the same set of RHLFs discussed
in Secs.6 generally agree with those derived from the NVSS at the limit of this
survey and within $z=0.2$ (Fig.\ref{conteggi_z02}). The flattening of the 
counts below $\sim 50-60$ mJy is both due to the combination of the 
low power cut-offs of the RHLFs with the redshift limit, and to the 
RHLFs evolution with redshift.
On the other hand, past calculations which assume a fixed 
fraction of GRHs with cluster mass predict 
an increasing number of sources at lower fluxes
(e.g., En\ss lin \& R\"ottgering, 2002).

GRHs around the peak of our LFs ($P_{1.4 GHz} \sim 10^{24}$W/Hz) and 
at z$\sim$0.15 would be detectable at fluxes below about 20 mJy,
which however is below the sensitivity limit of the NVSS for this
type of objects.
We estimate that the number of GRHs below this flux could be
up to 30-40 (whole sky, $z\le 0.2$) if superlinear scalings
of the mass with B hold.

The predicted number of GRHs (Fig.\ref{conteggi_z06_f})
(whole Universe) could be up to $\gtsim 100$ if a superlinear
scaling of the mass with B holds, while a sublinear scaling
would give a number 2-3 times smaller. 
A substantial number of these objects would be
found also down to a flux of a few mJy at 1.4 GHz in the case of
a superlinear scaling, while in the case of sublinear scalings
the number of GRHs below about 10 mJy would be negligible.

We also find that 
the bulk of GRHs is expected at $z \sim$0.1--0.3 (Fig.\ref{conteggi_z06}).
It should be mainly composed by those RHs populating the 
peak of the RHLFs, i.e. objects similar (or slightly more
powerful) to the GRH in the Coma cluster.\\

\item[$\bullet$]{\it Toward expectations at low radio frequencies}
\\

In Sec.7 we have extended our estimates to the case of low
frequency observations which will be made with upcoming instruments,
such as LOFAR and LWA. Lower energetic electrons contribute to these frequencies
and thus - in the framework of the particle re-acceleration scenario -
the efficiency of producing GRHs in galaxy clusters is expected to be
higher than that of GRHs emitting at 1.4 GHz.

By presenting the analysis for a representative set of parameters,
we have shown that the probability to have GRHs emitting at 150 MHz is 
significantly larger than that of those emitting at 1.4 GHz,
 particularly in the mass range $\sim\,5\cdot10^{14}-1.5\cdot10^{15}\,M_{\odot}$. 
 Consequently, the low mass cut-off in the RHMFs is shifted 
 down by a factor of $\sim$ 2.
This is naturally expected and is due to the fact that slightly
less turbulent systems are able to generate GRHs at lower frequencies.

We have also estimated that the number counts of GRHs at low frequencies
might outnumber those at 1.4 GHz by at least one order of magnitude. We venture
to predict that LOFAR is likely to discover $\gtsim 10^3$ (all sky) 
GRHs down to a flux of few mJy at 150 MHz.

\end{itemize}

\section*{Acknowledgments}
RC warmly thank S.Ettori and C.Lari for useful discussions
on the statistical analysis.
We acknowledge D.Dallacasa, K.Dolag and L.Feretti for comments
on the manuscript, and T.Ensslin for useful
discussions and for kindly providing the 
Local RHLFs in Fig.~\ref{RHLF_noi_ens}. The anonymous referee is
acknowledged for useful comments. RC acknowledge the MPA in 
Garching for the hospitality during the preparation of this paper.
This work is partially supported by MIUR under grant PRIN2004.

\label{lastpage}


\begin{thebibliography}{}

\bibitem{} Akritas M.G.,Bershady M.A., 1996, ApJ 470, 706
\bibitem{} Bacchi M., Feretti L., Giovannini G., Govoni F., 2003, A\&A 400, 465

\bibitem{} Bertoglio J.-P.,Bataille F., Marion J.-D., 2001, Physics of Fluids 13, 290

\bibitem{} Berezinsky V.S., Blasi P., Ptuskin V.S., 1997,
ApJ 487, 529
\bibitem{} B\"ohringer H., Schuecker P., Guzzo L., Collins C. A., Voges W., 
Cruddace R. G., Ortiz-Gil A., Chincarini G., De Grandi S., Edge A. C., and 4 coauthors,
2004, A\&A 425, 367
\bibitem{} Blasi P., 2004, JKAS 37, 483.
\bibitem{} Bowyer S., Korpela E.J., Lampton M., Jones T.W., 2004, ApJ 605, 168.
\bibitem{} Brunetti G., 2003, in 'Matter and Energy in Clusters of Galaxis', ASP Conf. Series, vol.301, p.349, 
eds. S. Bowyer and C.-Y. Hwang 
\bibitem{} Brunetti G., 2004, JKAS 37, 493
\bibitem{} Brunetti G., 2006, Astronomische Nachrichten 327, No 5/6, 615 
(proceedings of "The Origin and Evolution of Cosmic Magnetism", 29 August - 2 September 2005, Bologna, Italy)
\bibitem{} Brunetti G., Setti G., Feretti L., Giovannini G., 2001, MNRAS 320, 365
\bibitem{} Brunetti G., Blasi P., Cassano R., Gabici S., 2004, MNRAS 350, 1174
\bibitem{} Brunetti G., Blasi P., 2005, MNRAS 363, 1173
\bibitem{} Buote D.A, 2001, ApJ 553, 15
\bibitem{} Carroll S.M., Press W.H., Turner E.L., 1992, ARA\&A 30, 499
\bibitem{} Cassano R. \& Brunetti G., 2005, MNRAS 357, 1313

\bibitem{} Colafrancesco S., 1999, in "Diffuse thermal and relativistic plasma in galaxy clusters". Edited by Bohringer H., Feretti L., Schuecker P.. Garching, Germany : Max-Planck-Institut fur Extraterrestrische Physik, 1999. ("Proceedings of the Workshop...Ringberg Castle, Germany, April 19-23, 1999".), p.269
\bibitem{} Colafrancesco S., Marchegiani P., Perola G.C., 2005b, A\&A in press, astro-ph/0506565

\bibitem{} Condon J.J., Cotton W.D., Greisen E.W., Yin Q.F., Perley R.A., Taylor G.B., Broderick J.J.,
 1998, AJ 115, 1693
 
\bibitem{} David L. P., Slyz A., Jones C., Forman W., Vrtilek S. D., Arnaud K. A.,
1993, ApJ 412, 479
\bibitem{} Dennison B., 1980, ApJ 239
\bibitem{} Deiss B.M., Reich W., Lesch H., Wielebinski R., 1997 A\&A 321, 55

\bibitem{} Dolag K., 2006, invited review, Astronomische Nachrichten in press (proceedings of "The Origin and Evolution of Cosmic Magnetism", 29 August - 2 September 2005, Bologna, Italy), astro-ph/0601484
\bibitem{} Dolag K., Bartelmann M., Lesch H, 2002, A\&A 387, 383
\bibitem{} Dolag K., Grasso D., Springel V., Tkachev I., 2004, JKAS 37, 427

\bibitem{} Ebeling H., Voges W., Bohringer H., Edge A. C., Huchra J. P., 
Briel U. G., 1996, MNRAS 281, 799
\bibitem{} Ebeling H., Edge A. C., Bohringer H., Allen S. W., Crawford C. S., Fabian A. C., Voges W., Huchra J. P., 1998, MNRAS 301, 881

\bibitem{} En\ss lin T.A., 2004, 2004, JKAS 37, 439
\bibitem{} En\ss lin T.A., Biermann P.L.; Kronberg P.P., Wu X.-P., 1997, ApJ 477, 560
\bibitem{} En\ss lin T.A., Biermann P.L., Klein U.; Kohle S., 1998, A\&A 332 395
\bibitem{} En\ss lin T.A., Gopal-Krishna,  2001, A\&A 366, 26
\bibitem{} En\ss lin T.A., R\"ottgering H., 2002, A\&A, 396, 83


\bibitem{} Evrard A.E., Metzler C.A., Navarro J.F., 1996, ApJ 469, 494

\bibitem{} Ettori S., Fabian A.C., 1999, MNRAS 305, 834
\bibitem{} Ettori S., Tozzi P., Borgani S., Rosati P., 2004 A\&A 417, 13

\bibitem{} Feretti L., 2000, Invited review at IAU 199 `The Universe at Low Radio Frequencies' in Pune, India, 1999
\bibitem{} Feretti L., 2002, in "The Universe at Low Radio Frequencies", Proceedings of IAU Symposium 199, held 30 Nov - 4 Dec 1999, Pune, India. Edited by A. Pramesh Rao, G. Swarup, and Gopal-Krishna, p.133
\bibitem{} Feretti L., 2003, in "Matter and Energy in Clusters of Galaxis", ASP Conf. Series, vol.301, p.143, eds. S. Bowyer and C.-Y. Hwang.
\bibitem{} Feretti L., 2004, in "X-Ray and Radio Connections", eds. L.O. Sjouwerman and K.K. Dyer,
published electronically by NRAO, Held 3-6 Febraury 2004 in Santa Fe, New Mexico, USA. 
\bibitem{} Feretti L., Fusco-Femiano R., Giovannini G., Govoni F, 2001, A\&A 373, 106
\bibitem{} Feretti L., Brunetti G., Giovannini G., Kassim N., Orr\'u E., Setti G., 2004, JKAS 37, 315
\bibitem{} Feretti L., Orr\'u E., Brunetti G., Giovannini G., Kassim N., Setti G., 2004, A\&A 423, 111

\bibitem{} Fusco-Femiano R.,Dal Fiume D., Orlandini M., De Grandi S., 
Molendi S., Feretti L., Grandi P., Giovannini G., 2003, in `Matter and 
Energy in Clusters of Galaxis', ASP Conf. Series, eds., vol.301, p.109, 
S. Bowyer and C.-Y. Hwang

\bibitem{} Fusco-Femiano R., Orlandini M., Brunetti G., Feretti L., Giovannini G., Grandi P., Setti G., 2004, ApJ 602, 73
\bibitem{} Fujita Y., Takizawa M., Sarazin C.L., 2003, 
ApJ 584, 190
\bibitem{} Gabici S., Balsi P., 2003, ApJ 583, 695
\bibitem{} Giovannini G., Tordi M., Feretti L., 1999, NewA 4, 141
\bibitem{} Giovannini G., Feretti L., 2000, NewA 5, 335
\bibitem{} Giovannini G., Feretti L., 2002 in 'Merging Processes in Galaxy Cluster', vol.272, p.197, eds. L.Feretti, I.M.Gioia, G.Giovannini

\bibitem{} Govoni F., Feretti L., Giovannini G., B\"oringer H., Reiprich T.H., Murgia M., 2001, A\&A, 376, 803
\bibitem{} Govoni F., Markevitch M., Vikhlinin A., VanSpeybroeck L., Feretti, L., Giovannini G., 2004, ApJ 605, 695
\bibitem{} Govoni F., Feretti L., 2004, Int. J. Mod. Phys. D 13, 1549
\bibitem{} Govoni F., Murgia M., Feretti L., Giovannini G., Dallacasa D., Taylor G. B., 2005, A\&A 430, 5
\bibitem{} Hwang C.-Y., 2004, JKAS 37, 461
\bibitem{} Hoeft M., Brüggen M.; Yepes G., 2004, MNRAS 347, 389
\bibitem{} Hogg D.W., 2000, astro-ph/9905116
\bibitem{} Hughes J.P., Butcher J.A., Stewart G.C., Tanaka Y., 1993, ApJ 404, 611
\bibitem{} Kaastra J.S., Lieu R., Tamura T., Paerels F.B.S., den Herder J.W., 2003, A\&A 397 445
\bibitem{} Kempner J.C., Sarazin C.L., 2001, ApJ 548, 639
\bibitem{} Kim K.-T., Kronberg P.P., Dewdney P.E., Landecker T.L, 1990, ApJ 355, 29
\bibitem{} Kitayama T., Suto Y., 1996, ApJ.469, 480
\bibitem{} Kuo P.-H., Hwang C.-Y., Ip W.-H., 2004, ApJ 604, 108
\bibitem{} Kuo P.-H., Hwang C.-Y.; Ip, W.-H., 2003, ApJ 594, 732

\bibitem{} Lacey, C., Cole S., 1993, MNRAS 262, 627

\bibitem{} Lazarian A., 2006, invited review, Astronomische Nachrichten in press (proceedings of "The Origin and Evolution of Cosmic Magnetism", 29 August - 2 September 2005, Bologna, Italy)

\bibitem{} Lemonon L., Pierre M., Hunstead R., Reid A., Mellier Y., Boehringer H., 1997, A\&A 326, 34

\bibitem{} Liang H., 1999, in "Diffuse thermal and relativistic plasma in galaxy clusters". Edited by Bohringer H., Feretti L., Schuecker P.. Garching, Germany : Max-Planck-Institut fur Extraterrestrische Physik, 1999. ("Proceedings of the Workshop...Ringberg Castle, Germany, April 19-23, 1999".), p.33
\bibitem{} Liang H., Hunstead R.W., Birkinshaw M., Andreani P., 2000, ApJ 544, 686

\bibitem{} Markevitch M., 1996, ApJ, 465, 1
\bibitem{} Markevitch M., Forman W. R., Sarazin C. L., Vikhlinin A., 1998, ApJ 503, 77
\bibitem{} Markevitch M., Gonzalez A.H., David L., Vikhlinin A., Murray S., Forman W., Jones C., Tucker W.,
2002, ApJ 567, 27
\bibitem{} Miniati F., Jones T.W., Kang H., Ryu D., 2001, ApJ 562, 233
\bibitem{} Mushotzky R. F., Scharf C. A.,1997, ApJ 482, 13
\bibitem{} Nevalainen J., Markevitch M., Forman W., 2000, ApJ 532, 694
\bibitem{} Petrosian V., 2001, ApJ 557, 560 
\bibitem{} Pfrommer C., En\ss lin T.A., 2004, A\&A 413, 17

\bibitem{} Pierre M., Matsumoto H., Tsuru T., Ebeling H., Hunstead R., 1999, A\&AS 136, 173
\bibitem{} Press W.H., Schechter P., 1974, ApJ 187, 425
\bibitem{} Rephaeli Y., Gruber D., 2003, ApJ 595, 137
\bibitem{} Reimer O., Pohl M., Sreekumar P., Mattox J.R., 2003, ApJ 588,155
\bibitem{} Reimer A., Reimer O., Schlickeiser R., Iyudin A., 2004, A\&A 424, 773
\bibitem{} Reiprich T.H., B\"ohringer H., 2002, ApJ 567, 716
\bibitem{} Roettiger K., Burns J.O., Loken C., 1996, ApJ 473, 651
\bibitem{} Roettiger K., Burns J.O., Stone J.M., 1999, ApJ 518, 603
\bibitem{} Ryu D., Kang H., Hallman E., Jones T.W., 2003, ApJ 593, 599

\bibitem{} Sarazin C.L., 1986, Reviews of Modern Physics 58, 1
\bibitem{} Sarazin C.L., 1999, ApJ 520, 529
bibitem{} Sarazin C.L., 2002, in `Merging Processes in Clusters of Galaxies',
 vol.272, p.1, edited by L. Feretti, I. M. Gioia, and G. Giovannini
\bibitem{} Schindler S., 1996, A\&A 305, 756	
\bibitem{} Schindler S., 2002, in "Merging Processes in Galaxy Clusters." Edited by L. Feretti, I.M. Gioia, G. Giovannini. Astrophysics and Space Science Library, Vol. 272. Kluwer Academic Publishers, Dordrecht, 2002, p. 229-251
\bibitem{} Schuecker P., B\"ohringer H.; Reiprich T.H., Feretti L., 2001, A\&A 378, 408
\bibitem{} Schlickeiser R., Sievers A., Thiemann H., 1987, A\&A 182, 21

\bibitem{} Tsuru T., Koyama K., Hughes J.P., Arimito N., Kii T., Hattori M., 1996, in 
"UV and X-ray Spectroscopy of Astrophysical and Laboratory Plasmas": Edited by K. Yamashita and T. Watanabe. Tokyo : Universal Academy Press, 1996. (Frontiers science series ; no. 15)., p.375
\bibitem{} Vazza F., Tormen G., Cassano R., Brunetti G., Dolag K., 2006,
accepted for publication in MNRAS Letters, astro-ph/0602247 
\bibitem{} Venturi T., Bardelli S., Dallacasa D., Brunetti G., Giacintucci S., Hunstead R.W., Morganti R., 2003, A\&A 402, 913
\bibitem{} V\"{o}lk H.J., Aharonian F.A., Breitschwerdt D., 1996,
SSRv 75, 279
\bibitem{} White D. A., 2000, MNRAS 312, 663
\bibitem{} Zhang Y.-Y., Finoguenov A., Böhringer H., Ikebe Y., Matsushita K.,
Schuecker P., 2004, A\&A 413, 49

\end{thebibliography}
\end{document}